\documentclass[11pt]{article}

\usepackage{amsmath,amsfonts,amssymb,amsbsy,textcomp}
\usepackage{graphicx,color}
\usepackage{graphics}

\usepackage{amsmath, amssymb,graphicx}
\usepackage{epsfig}
\usepackage{verbatim}
\usepackage{amsfonts}
\usepackage{cite}

\usepackage{epsf}

\usepackage{tikz}
\usetikzlibrary{arrows,shapes}

\newcommand{\half}{\frac{1}{2}}

\newcommand{\Smat}{{\mathcal S}}
\newcommand{\Refl}{{\mathcal R}}

\newcommand{\8}{{\infty}}

\newcommand{\eps}{\epsilon}
\newcommand{\tr}{\,{\rm tr}}

\newcommand{\bra}[1]{{\,\langle#1|}\,}
\newcommand{\ket}[1]{{\,|#1\rangle }\,}

\newcommand{\C}{\mathcal C}

\newcommand{\dby}[1]{\frac{\partial}{\partial #1}}

\newcommand{\btp}{\begin{tikzpicture}[baseline=0pt,scale=0.9,line width=0.7pt]}
\newcommand{\etp}{\end{tikzpicture}}

\def\be{\begin{equation}}
\def\ee{\end{equation}}
\def\ba{\begin{eqnarray}}
\def\ea{\end{eqnarray}}
\def\bc{\begin{center}}
\def\ec{\end{center}}
\def\cO{{\mathcal{O}}}

\def\nn{\nonumber}

\def\r2{{\sqrt{2}}}

\def\msu{{\mathfrak{su}}}

\newcommand{\bea}{\begin{eqnarray}\begin{array}}
\newcommand{\eea}{\end{array}\end{eqnarray}}

\begin{document}

\begin{titlepage}
\begin{flushright}
{\bf \today} \\
DAMTP-2009-39\\
DCPT-09/29\\
arXiv:0905.1700 [hep-th] \\
\end{flushright}
\begin{centering}
\vspace{.2in}

 {\Large {\bf
Finite size corrections for open strings/open chains \\
 in planar AdS/CFT
}}

\vspace{.3in}

{\large D. H. Correa ${}^{a,1}$ and C. A. S. Young ${}^{b,2}$}\\
\vspace{.2 in}
${}^{a}${\emph{DAMTP, Centre for Mathematical Sciences \\
University of Cambridge\\ Wilberforce Road,
Cambridge CB3 0WA, UK}} \\
\vspace{.2in}
\vspace{.1 in}
${}^{b}${\emph{Department of Mathematical Sciences\\ University of Durham\\
South Road, Durham DH1 3LE, UK}}

%
\footnotetext[1]{{\tt D.Correa@damtp.cam.ac.uk,}\quad ${}^{2}${\tt charles.young@durham.ac.uk}}
\vspace{.5in}

{\bf Abstract}

\vspace{.1in}

\end{centering}

We identify the leading finite-size (L\"uscher-type) correction to the energy of open strings ending on maximal giant gravitons. In particular we  obtain the leading finite size correction at weak 't Hooft coupling and in the planar limit to the energy of very short vacuum states. These results are shown to agree with certain 1, 2, 3 and 4-loop dual gauge theory perturbative calculations, which we also perform.

\end{titlepage}

\tableofcontents

\section{Introduction}
As a result of much progress in recent years (see e.g. \cite{MZ,BS,BKS,AFS,BKSZ}) using the methods of integrable systems, the spectrum of anomalous dimensions of single-trace operators in planar $\mathcal N=4$ super Yang-Mills theory can now be computed exactly, for any value of the 't Hooft coupling, in the limit in which these operators are very long. It is believed that this spectrum, or equivalently (on the string side of the AdS/CFT correspondence \cite{adscftmalda}) that of the energies of free strings in $AdS_5\times S^5$ with large angular momentum, is fully determined by a certain system of asymptotic Bethe ansatz (ABA) equations \cite{BSpsu224}.

Much recent effort \cite{wrappingwork}  has therefore focussed on extending these results to operators of finite length (meaning traces of finitely many fields) where ``wrapping effects'' \cite{BDS} not captured by the ABA \cite{AJK} must be taken into account. It now appears that the correct framework for performing these computations is that of the thermodynamic Bethe ansatz (TBA) \cite{TBA}, which allows physical quantities for a system of finite size to be extracted from the infinite-volume data (specifically, from the asymptotic S-matrix \cite{Smat}). A most remarkable result, obtained in the last year, was the computation, using methods based on TBA ideas and on the work of L\"uscher \cite{Luscher}, of the leading weak-coupling finite-size correction to the anomalous dimension of the Konishi operator \cite{BJ}. This occurs at four-loop order in the 't Hooft coupling and involves an intricate sum of rational and transcendental pieces. When this correction is included, the complete answer for the anomalous dimension matches that obtained by the (technically daunting) direct gauge-theory calculation \cite{FSSZandV}. This is strong evidence that the TBA approach is correct. More recently, a string hypothesis has been formulated for the mirror model \cite{Arutyunov:2009zu} and the TBA equations and associated Y-system for AdS/CFT have been proposed \cite{GKV,BFT,GKKV,AFtba}.

In the present work we begin the study of finite-size effects for operators with \emph{boundaries}, or equivalently, for \emph{open} strings. In various setups, open strings are integrable classically \cite{Mann} and in the weak coupling limit \cite{BV}. In those cases integrability is believed to hold at all values of the coupling \cite{HMopen,Galleas}. There are various motivations for considering the question of open-boundary finite-size effects. In the first place, open strings are part of the spectrum of the theory and eventually one would like to have the tools to describe them. To this end, it is indispensable  to incorporate the finite size corrections to the open-boundaries asymptotic Bethe ansatz. The boundary version of the thermodynamic Bethe ansatz (BTBA) \cite{LMSS} should be the framework for doing so. As we recall below, the general structure of BTBA equations can be used to extract the form of the leading L\"uscher-type correction to the ground state energy for the worldsheet QFT defined in a strip of finite width $L$. In this article, we will compute such leading finite size corrections.

A second, and maybe more important motivation, is that the open boundaries setup will provide an excellent laboratory, where a variety of simple calculations can be carried through to test the correctness of the TBA method as the tool for incorporating finite size corrections in the planar AdS/CFT spectrum. To begin with, there \emph{is} a non-trivial ground-state BTBA calculation to be done, in contrast to the closed case, because the Bethe vacuum state is no longer protected by supersymmetry -- and, though there are various approaches to computing excited-state energies \cite{exciTBA}, it is always the ground-state energy which emerges most directly from TBA methods. Moreover, as we shall discuss, the leading finite-size effects can appear as early as 1- or 2-loop order in weak-coupling perturbation theory. For such cases it will be easy to perform explicit computations in the dual $\mathcal N=4$ SYM gauge theory to compare with the worldsheet QFT.

Another interesting feature of the open boundaries setup is the following interplay between alternative reference vacua. When using a Bethe ansatz to describe the spectrum of open strings attached to giant gravitons (spherical D3-branes
carrying angular momentum), there are two physically inequivalent possibilities: either both the D-brane and the reference  state carry angular momentum in the same direction, or they do not. In the dual conformal field theory this translates to using the same, or two different, scalar fields to represent the D-brane and the reference state. Now, if a few impurities are added to a very short reference state, we can swap the roles between impurities and background fields. Consider for instance the operator
\begin{equation}
\cO_{Z}(YZY)
\equiv\epsilon^{i_1,\cdots, i_N}_{j_1,\cdots ,j_N}Z^{j_1}_{i_1} \cdots
Z^{j_{N-1}}_{i_{N-1}}(YZY)^{j_N}_{i_N}\,.
\label{example}
\end{equation}
One can choose to regard this as a state with boundary impurities $Y$ in a background of scalar fields $Z$ or, alternatively, as a state with a bulk impurity $Z$ in a background of scalar fields $Y$. The asymptotic Bethe ans\"atze for these possibilities will not give the same answer for the anomalous dimension, and, of course, neither of them will give the  correct finite-volume anomalous dimension. Each will nevertheless capture the  finite-volume anomalous dimension up to certain order in the perturbative weak-coupling expansion. And, interestingly, one of the points of view will be more efficient,   in the sense that it will capture the exact anomalous dimension to a higher loop order than the alternative point of view.

This constitutes a potentially powerful tool for producing tests of BTBA results, without performing explicit perturbative calculations: one can simply use the asymptotic Bethe ansatz answer of the more efficient point of view to test the finite size corrected answer of the less efficient point of view. It should be emphasized that this interplay between  alternative points of view is an attribute of the open boundaries cases exclusively. One could also swap the roles between impurities and background fields in a short single trace. However, the asymptotic Bethe ansatz and the finite size corrected answers would be essentially the same.

This paper is structured as follows: after quickly listing our conventions, we recall in section 2 some details of the boundary states setup and the boundary L\"uscher corrections. Then in sections 3 and 4 we apply this method to compute the corrections to the energies of strings ending on ``$Y=0$'' and ``$Z=0$'' (in the sense of \cite{HMopen}) maximal giant gravitons.
We conclude with some comments on future possibilities in section 5. The reflection matrix for $Q$-magnon bound states is given in an appendix.

\subsection{Notation and Conventions}
The idea behind the TBA approach is that the partition function of an integrable 2-dimensional QFT may be evaluated either in the original (physical) theory or in the mirror theory. The latter is obtained through a double Wick rotation that takes $p\mapsto iE$ and $E\mapsto ip$.

Having this in mind, let us quickly set up  our conventions to characterize particles in both the physical and the mirror models. Bound states of $Q\in \{1,2,\dots\}$ magnons are described by the spectral parameters $x^{\pm}$, which satisfy the mass-shell condition
\be
x^+ +\frac1{x^+}-x^--\frac1{x^-} = \frac{iQ}{g}\,,
\ee
where $g$ is related to the 't Hooft coupling of the gauge theory by $\lambda = {16\pi^2} g^2$. The momentum $p$ and energy $E$ of a physical bound state of magnons are given by
\be
e^{ip}=\frac{x^+}{x^-}\,,\qquad E = Q + 2ig \left(\frac{1}{x^+}-\frac{1}{x^-}\right),
\ee
and they satisfy the physical dispersion relation
\be
E^2 = Q^2 + 16 g^2 \sin(\tfrac{p}2)^2\,.
\ee
Alternatively, bound states of magnons can be described by a generalized rapidity $\zeta$ using Jacobi elliptic functions:
\be
x^{\pm}(\zeta) =
\frac{Q}{4g} \left(\frac{{\rm cn}(\zeta,k^2)}{{\rm sn}(\zeta,k^2)}\pm i\right)
 (1+{\rm dn}(\zeta,k^2))\,.
\label{Defellipxpxm}
\ee
The momentum $p(\zeta)$ and energy $E(\zeta)$ obey
\be
p(\zeta)=2{\rm am}(\zeta,k^2)\,,~~ \sin(\tfrac{p(\zeta)}{2})={\rm sn}(\zeta,k^2)\,,~~
     E(\zeta)= Q {\rm dn}(\zeta,k^2)\,,\label{DefellippE}
\ee
Here the elliptic modulus $k^2=-\frac{16g^2}{Q^2}$ is real for real values of the coupling $g$.
The rapidity $\zeta$ takes values on the
complex torus defined by $\zeta\sim \zeta+ 2\omega_1\sim\zeta+ 2\omega_2$, with the half-periods
\be
\omega_1 = 2 {\rm K}(k^2)\,,\qquad \omega_2 = 2 i{\rm K}(1-k^2)-2 {\rm K}(k^2)\,,
\label{EllipPeriods}
\ee
where ${\rm K}(k^2)$ is the complete elliptic integral of the first
kind. For real $g$, $\omega_1\in \mathbb R$ and $\omega_2\in i\mathbb R$.

The shift $\zeta \mapsto \zeta \pm \omega_1$ leaves $x^\pm(\zeta)$ invariant. But sending
$\zeta\mapsto \zeta \pm \omega_2$ performs a crossing transformation between the forward and backward
 mass-shells:
\be
x^\pm (\zeta\pm\omega_2)
=\frac{1}{x^{\pm}(\zeta)}\,,\label{xcrossingTrans}
\quad E(\zeta\pm \omega_2) = - E(\zeta)\,, \quad p(\zeta\pm\omega_2) = -p(\zeta)\,.
\ee

The double Wick rotation that will take us to the mirror theory in the TBA approach can also be implemented through a shift of the physical rapidity \cite{AF2}. Thus, we introduce the mirror rapidity $\tilde\zeta$, which we take to be
\be
\tilde\zeta = \zeta + \frac{\omega_2}{2}\,.
\ee
The mirror momentum $q$ and energy $\tilde E$, are defined by
\be
q = iE = \sqrt{1-k^2} Q\, {\rm sc} (\tilde\zeta,k^2) ,\quad
\tilde E = ip = 2i\arcsin\left(\frac{Q}{4ig}\,{\rm dc}(\tilde\zeta,k^2)\right) .
\label{EQrel}\ee
The particle is on-shell for all values of $\zeta$, or $\tilde\zeta$, but it has \emph{real} energy
and momentum only for real $\zeta$, and real mirror energy and momentum only for real $\tilde\zeta$.
Thus, physical particles have real $\zeta$ and mirror particles have real $\tilde\zeta$. By convention, we shall write the spectral parameters of mirror particles as $z^\pm$, reserving $x^\pm$ for physical particles.

Note that, in contrast to relativistic theories, physical and mirror particles possess different kinematics. The mirror dispersion relation reads
\be
\sinh(\tfrac{\tilde{E}}{2})^2 = \frac{Q^2+q^2}{16g^2} \, .
\ee
Also, on the real interval $(-\frac{\omega_1}{2},\frac{\omega_1}{2})$ the function $p(\zeta)\in \mathbb R$ is odd and increases monotonically, with
\be
p(\zeta) \to \pm \pi \quad\text{as}\quad \zeta \to \pm\frac{\omega_1}{2}\,,
\label{allp}
\ee
while mirror magnon momentum can take any real value
\be
q(\tilde\zeta) \to \pm \8 \quad\text{as}\quad \tilde\zeta \to \pm\frac{\omega_1}{2}\,.
\label{allpm}
\ee

\section{Boundary States, L\"uscher corrections and Boundary TBA}
\label{BSBTBA}
Let us recall some details of the boundary state formalism \cite{GZ}, boundary L\"uscher corrections and the boundary thermodynamic Bethe ansatz equation \cite{LMSS,DPTW}. We write $L$ for the system size, i.e. the distance between the left and right boundaries (the units of $L$ and its interpretation in the AdS/CFT context are discussed below). Consider compactifying the Euclidean time direction of the system on a circle of circumference $R$. Then, as usual in TBA approaches, the idea is that the partition function $Z(L,R)$ may be evaluated in two different ways (figure 1). On the one hand,
\be Z(L,R) = \tr_{\mathcal H_P(L)} e^{-R H_P(L)} \ee
where $\mathcal H_P(L)$ and $H_P(L)$ are the Hilbert space and Hamiltonian of the original, physical, theory. They depend on the system size, and, in the present case, also encode the details of the boundary conditions. In the limit $R\rightarrow \8$,
\be Z(L,R) \sim e^{-R({\cal E}_0(L)-{\cal E}_0(\8))},\ee
where ${\cal E}_0(L)$ is the energy of the lowest lying state of the system as a function of $L$, which is the quantity we would like to compute.

Alternatively, we can regard the system as evolving in the Euclidean time direction of the mirror theory (which we think of as running from right to left).
Provided $R$ is very large, we may take the time evolution operator $H_M$ and space of states $\mathcal H_M$
to be simply those of the mirror theory in infinite volume. What were the right and left
boundaries now correspond to, respectively, the initial and final {\it states}, and the partition function is thus of the form

\begin{figure}
\begin{center}
\begin{tabular}[3]{ccc}
\begin{tikzpicture}[x={(0cm,1cm)},y={(.25cm,0cm)}]
\draw[very thick] (0,-3) circle (2) (0,3) circle (2);
\fill[white,fill=white] (-2,-3) -- (2,-3) -- (2,0) -- (-2,0);
\draw[very thick] (-2,-3) -- (-2,3);
\draw[|<->|] (2.2,-3) -- (2.2,3) node [midway,above] {$L$};
\draw[very thick] (2,-3) -- (2,3);
\draw[dashed] (0,-3) circle (2);
\draw[<->] (2,-4) arc (-0:-180:2);
\node at (0,-7) {$R$};
\end{tikzpicture} &$\qquad\qquad$&
\begin{tikzpicture}[x=1cm,y=.25cm]
\draw[very thick] (0,-3) circle (2) (0,3) circle (2);
\fill[white,fill=white] (-2,-3) -- (2,-3) -- (2,0) -- (-2,0);
\draw[very thick] (-2,-3) -- (-2,3);
\draw[|<->|] (2.2,-3) -- (2.2,3) node [midway,right] {$L$};
\draw[very thick] (2,-3) -- (2,3);
\draw[dashed] (0,-3) circle (2);
\draw[<->] (2,-4) arc (-0:-180:2);
\node at (0,-7) {$R$};
\node at (0,-3) {$\ket{B^{\text {right}}}$};
\node at (0,3) {$\ket{B^{\text{left}}}$};
\end{tikzpicture} \\
\qquad Physical && \!\!\!\!\!\!\!\!\!\!\!\! Mirror
\end{tabular}
\end{center}
\caption{Alternative viewpoints for the partition function $Z(L,R)$} \label{fig0}
\end{figure}
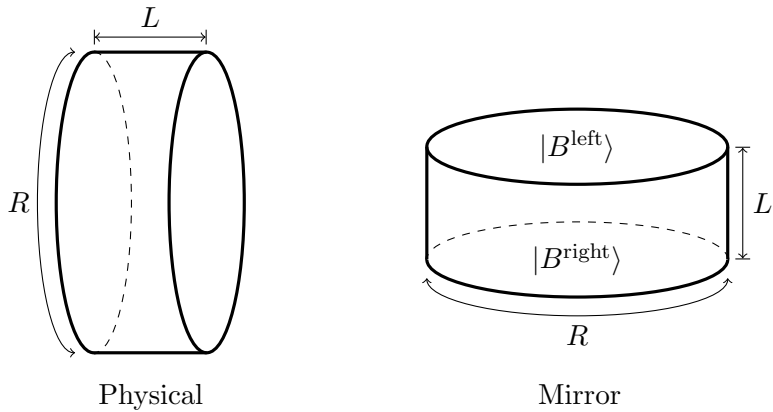

\be
Z(L,R) = \bra{B^{{\text{left}}}} e^{-L H_M} \ket{B^{{\text{right}}}}.
\ee
Such boundary states $\ket B$ were first introduced by Ghoshal and Zamolodchikov \cite{GZ}. For a discussion of their appearance in the present context see \cite{Palla}. The expansion of the initial/right boundary state in the basis of scattering out-states of the mirror theory is written
\be
\ket{B^{\text{right}}} = \ket{0}+\int_0^{\tfrac{\omega_1}2}\!d\tilde\zeta\,
 \ket{r,-\tilde\zeta; t,\tilde\zeta} K^{rt}(\tilde\zeta) + \dots\ee
where $\cdots$ represent terms with more particles.\footnote{If the boundary has a one-particle interaction, $\ket B$ may also have overlap with a one-particle state of zero rapidity. One-particle interactions are discussed below.}
Intuitively, $\ket B$ can be thought of as encoding all possible scattering processes of particles against the boundary. In particular its overlap with the two-particle mirror out-state $\ket{r,-\tilde\zeta;t,\tilde\zeta}$ ($\tilde\zeta>0$) is proportional to the amplitude for the reflection process $\bar r \to t$ at (non-real) rapidity
${ -\tilde\zeta+\frac{\omega_2}{2}}$ in the physical theory:
\be\btp
     \draw[very thick] (3,-2) -- (3,2);
     \draw[->] (3,0) -- ++(-38:-3) node[left] {$t,\tilde\zeta-\tfrac{\omega_2}{2}$};
     \draw     (3,0) -- ++(38:-1)   node[above left] {$\bar r,-\tilde\zeta+\tfrac{\omega_2}{2}$} ;
     \draw[<-] (3,0)++(38:-1) -- ++(38:-.75);
     \draw[-o] (3,0)++(38:-1.25) -- ++(38:-1)  node[below right] {$\C^{r\bar r }$};
     \draw[->] (3,0)++(38:-2.25) -- ++(38:-.75) node[left] {$r,-\tilde\zeta-\tfrac{\omega_2}{2}$};
\etp\ee
Thus
\be K^{rt}(\tilde\zeta)
=  \C^{r\bar r} \Refl^{\rm R}{}_{\bar r}{}^t(-\tilde\zeta + \tfrac{\omega_2}{2})\label{KR}\qquad
\ee
where $\C$ is the charge conjugation matrix and $\Refl^{\rm R}(\zeta)$ is the reflection matrix for the {right}  boundary. The integral above is over all positive real mirror momenta (c.f. \ref{allpm}) and thus over all two-particle out-states with vanishing total mirror momentum.
It must also be possible to write the boundary state in terms of the basis of mirror \emph{in}-states,
and by convention this decomposition is written
\be \ket{B_2^\text{{right}}} = \int_0^{\tfrac{\omega_1}2}\! d\tilde\zeta\,
\ket{r,\tilde\zeta; t,-\tilde\zeta} K^{rt}(-\tilde\zeta).\label{inL}
\ee
The in- and out-bases are related by the mirror S-matrix according to
\be
\ket{ p, \tilde\zeta;q,\tilde\zeta'} = \Smat_{pq}^{tr}(\tilde\zeta,\tilde\zeta')
\ket{ r,\tilde\zeta'; t,\tilde\zeta},
\label{inout}
\ee
which holds both for $\tilde\zeta>\tilde\zeta'$ and $\tilde\zeta<\tilde\zeta'$ by virtue of unitarity of the S matrix,
\be
\Smat_{pq}^{tr}(\tilde\zeta,\tilde\zeta') \Smat_{rt}^{xy} (\tilde\zeta',\tilde\zeta) = \delta_p^y \delta_q^x.
\label{unitarity}
\ee
It follows that we must have
\be
\btp
\draw[-](2,2)--(2,2) node[left]{$K^{pq}(\tilde\zeta) \Smat_{pq}^{tr} (-\tilde\zeta,\tilde\zeta) = K^{rt}(-\tilde\zeta)$};
\etp
\qquad\btp
     \draw[very thick] (-1.5,1) -- (1.5,1);
     \draw[->] (.75,2) -- ++(0,.1); \draw[->] (-.75,2) -- ++(0,.1);
     \draw (0,1) .. controls ++(-1,1) .. node[left] {$p,-\tilde\zeta$} (0,3);
     \draw (0,1) .. controls ++(1,+1) .. node[right] {$q, \tilde\zeta$}  (0,3) ;
     \draw[->] (0,3) -- (1,4) node[above] {$t,-\tilde\zeta$};
     \draw[->] (0,3) -- (-1,4) node[above] {$r,\tilde\zeta$};\etp
\qquad \btp  \draw[-](2,2)--(2.4,2); \draw[-](2,1.85)--(2.4,1.85); \etp
\qquad \btp
     \draw[very thick] (-1.5,1) -- (1.5,1);
     \draw[->] (0,1) -- (1,2) node[above] {$t,-\tilde\zeta$};
     \draw[->] (0,1) -- (-1,2) node[above] {$r,\tilde\zeta$};\etp
     \nn\ee
When combined with the demand that (\ref{KR}) should hold for $\tilde\zeta<0$ as well as $\tilde\zeta>0$, this constitutes a constraint on the reflection matrix $\Refl(\zeta)$, called the Boundary Crossing-Unitarity condition. This condition plays a major part in constraining the overall scalar factor of the reflection matrices \cite{HMopen,ChCo,ABR,ANZF,Palla}.

The left  boundary, corresponding to the final state, is most naturally pictured in terms of mirror in-states.
\be
\bra{B_2^{\text{left}}} = \int_0^{\tfrac{\omega_1}2}\! d\tilde\zeta\,\, \Refl^\text{L}{}_p{}^{\bar q}
(\tilde\zeta - \tfrac{\omega_2}{2}) \C_{\bar q q} \bra{p,\tilde\zeta;q,-\tilde\zeta} .\ee

Thus, in view of (\ref{inL}) and the equation above, the overlap between the two-particle final state at
Euclidean time $L$ and the two-particle initial state is\footnote{The $\delta(0)$ singularity comes from
a $\delta(\tilde\zeta-\tilde\zeta')^2$ in the integral. As in the relativistic case \cite{LMSS}, it is best to change variables to momentum in order to regularize. Doing so gives rise to a Jacobian factor $\delta(\tilde\zeta-\tilde\zeta')^2 \sim \tfrac{R}{2\pi}\tfrac{dq}{d\tilde\zeta}\delta(\tilde\zeta-\tilde\zeta')$.}
\ba
\bra{B_2^\text{left}}e^{-LH_M}\ket{B_2^\text{right}} &\sim& \delta(0)
      \int_0^{\tfrac{\omega_1}2}\!  d\tilde\zeta \, e^{-2\tilde E(\tilde\zeta) L} \, \chi(\tilde\zeta)\nn
\\
&\sim& \frac{R}{2\pi}
      \int_0^{\tfrac{\omega_1}2}\! \frac{dq}{d\tilde\zeta} d\tilde\zeta \, e^{-2\tilde E(\tilde\zeta) L} \, \chi(\tilde\zeta)\,  = \frac{R}{2\pi}\int_0^{\infty} {dq}\, e^{-2\tilde E(q) L} \, \chi(q)\,,
\ea
where
\ba \chi(\tilde\zeta) &\!=\!& \Refl^\text{L}{}_p{}^{\bar q} (\tilde\zeta-\tfrac{\omega_2}{2})
                          \Refl^\text{R}{}_{\bar p}{}^q (\tilde\zeta+\tfrac{\omega_2}{2})
                  \C^{p \bar p} \C_{\bar q q} \nn\\
&\!=\!& \Refl^\text{L}{}_p{}^{\bar q} (\tilde\zeta-\tfrac{\omega_2}{2})
                          \Refl^\text{L}{}_{\bar p}{}^q (-\tilde\zeta-\tfrac{\omega_2}{2})
                  \C^{p \bar p} \C_{\bar q q}
                  \,= \,\Refl^\text{L}{}_p{}^{\bar q} (\zeta)
                          \Refl^\text{L}{}_{\bar p}{}^q (-\zeta-{\omega_2})
                  \C^{p \bar p} \C_{\bar q q}.
                  \label{chi}
                  \ea
In the second line we used parity ($\Refl^\text{R}(\zeta) = \Refl^{\rm L}(-\zeta)$).
This quantity $\chi(\tilde\zeta)$ is a key ingredient in the boundary TBA equation
and includes all the particles of the mirror theory. In particular, in our case,
this will involve an infinite sum over magnon bound-states. We will write $\chi_Q(q)$ for the contribution from the $Q$-magnon bound states, whose energy $\tilde E_Q(q)$ is given by eq. (\ref{EQrel}). Now, in the $R\to\infty$ limit
\be
e^{-R({\cal E}_0(L)-{\cal E}_0(\8))} \sim
1 +  \frac{R}{2\pi}  \sum_{Q=1}^\infty\int_0^{\infty} {dq}\, e^{-2\tilde E_Q(q) L} \, \chi_Q(q)+\cdots
\label{2par}
\ee
In many cases, this two-particle contribution to the partition function dominates the leading finite size correction to the ground state energy.

However, this is not so in the case of $Z=0$ giant gravitons we consider in section \ref{z=0}. Rather, it will turn out that the leading finite size correction comes from a one-particle interaction at the boundary. Such interactions are encoded in the pole structure of the reflection matrix, and, as we now recall, one way to calculate their form is from the general structure of boundary TBA equations \cite{Bajnok:2004tq}. In a boundary thermodynamic Bethe ansatz (BTBA), the exact finite size correction for the energy of the ground state is given by
\be
{\cal E}_0(L)-{\cal E}_0(\infty) =-\frac{1}{2\pi} \sum_{Q=1}^\infty \int_0^{\infty} dq \log(1+\chi_Q(q)e^{-\epsilon_Q(q)})\,
\label{btbaene},
\ee
which has to be fed with  a set of functions $\epsilon_Q(q)$, known as pseudo-energies, that (together with additional pseudo-energies the higher levels of nesting) solve a set of BTBA equations. We do not attempt to derive the set of BTBA equations for this system here. For the leading order corrections we are concerned with, let us note that the pseudoenergies of the physical particles have the following asymptotics in the limit of large $L$,
\be
\epsilon_Q(q) \simeq 2L \tilde{E}_Q(q)\, . \label{btba}
\ee
We will assume (based on the form of iterative approximations to solutions of the simpler BTBA equations of purely diagonal-scattering models \cite{LMSS}) that the correction to this equation is $\mathcal O(e^{-2L\tilde E_Q(q)})$.
Although in general (\ref{btba})  is regarded as an infrared, i.e. large $L$, asymptotic \cite{Bajnok:2004tq}, it is possible to regard it instead as a weak coupling limit, because the 't Hooft coupling dependence of the mirror bound state magnons is such that
\be
 e^{-2L\tilde{E}_Q(q)} \simeq \left(\frac{4g^2}{Q^2+q^2}\right)^{2L}  \ll 1
\qquad {\rm for\ \ } g^2\ll 1 \quad {\rm and} \quad any\ \  L.
\ee
Thus the leading finite size (i.e. L\"uscher-type) correction, which we will be using in this paper, is
\be
{\cal E}_0^{(0)}(L)-{\cal E}_0(\infty) =-\frac{1}{2\pi} \sum_{Q=1}^\infty\int_0^{\infty} dq \log(1+\chi_Q(q)e^{-2L \tilde{E}_Q(q)})\,.
\label{btbaene0}
\ee
and what we will mean by leading finite size correction is, more precisely, the leading order in the weak coupling expansion of (\ref{btbaene0}).

There are two essentially different possibilities for approximating this integral, depending whether or not the boundaries have one-particle interactions.
Such interactions are present when the reflection matrix has a simple pole at the imaginary rapidity $\zeta = \tfrac{\omega_2}{2}$ ($-\tfrac{\omega_2}{2}$) for a right (respectively, left) boundary.
If there are no such interactions, each $\chi_Q(q)e^{-2L \tilde{E}_Q(q)}$ is much smaller than 1 for all
values of the momentum $q$ and the leading order of (\ref{btbaene0}) is captured by
\be
{\cal E}_0^{(0)}(L)-{\cal E}_0(\infty) \simeq -\frac{1}{2\pi} \sum_{Q=1}^\infty \int_0^{\infty} dq \chi_Q(q)e^{-2L \tilde{E}_Q(q)}\,,
\label{btbaene0a}
\ee
which is exactly the 2-particle contribution in (\ref{2par}).

On the other hand, if there are such one-particle interactions with the boundaries for some value of $Q$,
$\chi_Q(q)$ will have a double pole  at $q=0$. The expression (\ref{btbaene0a}) for the leading finite size correction is still valid. However, $\log(1+\chi_Q(q)e^{-2L \tilde{E}_Q(q)})$ can no longer be approximated by $\chi_Q(q)e^{-2L \tilde{E}_Q(q)}$  as $q$ approaches 0. Let us consider a case in which
\be
\chi_Q(q)e^{-2L \tilde{E}_Q(q)}\sim \frac{C^2}{q^2}\qquad {\rm for}\quad q\to 0\,.
\ee
Following  \cite{Bajnok:2004tq}, we can simply re-write (\ref{btbaene0}) as
\be
{\cal E}_0(L)-{\cal E}_0(\infty) = -\frac{1}{2\pi}\int_0^{\infty} dq \log\left(1+\frac{C^2}{q^2}\right)
-\frac{1}{2\pi}\int_0^{\infty} dq \log\left(\frac{1+\chi_Q(q)e^{-2L \tilde{E}_Q(q)}}{1+\frac{C^2}{q^2}}\right)\,.
\label{btbaene0b}
\ee
The first integral can be exactly solved, and the second one is sub-leading. Therefore,
\be
{\cal E}_0(L)-{\cal E}_0(\infty) = -\frac{1}{2}|C|+{\cal O}(C^2)\,.
\label{btbaene0bb}
\ee
It is this equation which will generate all the non-trivial finite size corrections we will compute in the present work. To proceed, we only need to compute the functions $\chi_Q(q)$
for our cases of interest.

\section{$Y=0$ brane}
In this section we shall consider an open string carrying $J$ units of $Z$ charge and ending on a $Y=0$ maximal giant graviton (which is a D3-brane carrying angular momentum along the $Y$ direction \cite{HMopen,McGreevy:2000cw}).
The corresponding operator in the dual conformal field theory is
\begin{equation}
\cO_{Y}(Z^J)=\epsilon^{i_1,\cdots,i_N}_{j_1,\cdots, j_N}Y^{j_1}_{i_1}\cdots
Y^{j_{N-1}}_{i_{N-1}} (Z^J)^{j_N}_{i_N}\label{DefOy}\,.
\end{equation}
Perturbative computations show that the anomalous dimension of this operator vanishes
in the large $N$ limit \cite{Balasubramanian:2002sa}.
However, this is not a BPS operator and one should ask whether, when the range
of interaction exceeds (twice) the length of the vacuum state, a finite size correction could
not lift the vanishing energy of the vacuum state.

To search for finite size corrections in this model, using the BTBA in the worldsheet QFT,
we need the boundary reflection matrices for all asymptotic states in the mirror theory. These states are accommodated in an infinite sum of short multiplets of $su(2|2)^2$.The $su(2|2)$ commutation relations are\footnote{We are using fundamental representation indices $a,b,\dots =1,2$ and $\alpha,\beta,\dots =3,4$}.
\begin{align}
[ {\mathfrak{R}}^{a}_{~{b}}, \mathfrak{J}^{c}]
&=
\delta^{c}_{b} \mathfrak{J}^{a}-\tfrac12 \delta^{a}_{b} \mathfrak{J}^{c}\,,
\qquad\;
[ {\mathfrak{R}}^{a}_{~{b}}, \mathfrak{J}_{c}]
=-\delta_{c}^{a} \mathfrak{J}_{b}+\tfrac12 \delta^{a}_{b} \mathfrak{J}_{c}\,,\nn
\\
[\mathfrak{L}^\alpha_{~\beta},\mathfrak{J}^\gamma]
&=\delta^{\gamma}_\beta \mathfrak{J}^\alpha-\tfrac12\delta^{\alpha}_\beta \mathfrak{J}^\gamma\,,
\qquad
[\mathfrak{L}^\alpha_{~\beta},\mathfrak{J}_{\gamma}]
= -\delta_{\gamma}^\alpha \mathfrak{J}_\beta+\tfrac12\delta^{\alpha}_\beta \mathfrak{J}_{\gamma} \,,
\label{eq1}\nn
\\
\{\mathfrak{Q}^{\alpha}_{~a},\mathfrak{Q}^{\beta}_{~ b}\}
&= \epsilon^{\alpha\beta}\epsilon_{ab}\mathfrak{P} \,,\qquad \quad \
\{\mathfrak{S}^{a}_{~\alpha},\mathfrak{S}^{b}_{~ \beta}\} = \epsilon_{\alpha\beta}\epsilon^{ab}\mathfrak{K} \,,\nn
\\
\{\mathfrak{S}^{a}_{~\alpha},\mathfrak{Q}^{\beta}_{~ b}\}
&= \delta^{ a}_{ b} \mathfrak{L}^\beta_{~\alpha}
+  \delta^\beta_{\alpha}  {\mathfrak{R}}^{ a}_{~ b}
+ \delta^{ a}_{ b} \delta^\beta_{\alpha} \mathfrak{C} \,.\nn
\end{align}
The transformation rules for a magnon in the fundamental representation are given by
\bea{lll}
&\mathfrak{R}^a_{~b} |\phi^c\rangle  =
\delta^c_b |\phi^a\rangle -\frac{1}{2} \delta^a_b |\phi^c\rangle\,,
 &
\mathfrak{L}^\alpha_{~\beta}|\psi^\gamma\rangle =
\delta^\gamma_\beta |\psi^\alpha\rangle -\frac{1}{2} \delta^\alpha_\beta |\psi^\gamma\rangle\,,
\\
&\mathfrak{Q}^\alpha_{~a}|\phi^b\rangle =
a\ \delta^b_a |\psi^\alpha\rangle\,,
 &
\mathfrak{Q}^\alpha_{~a}|\psi^\beta\rangle =
b\ \epsilon^{\alpha\beta}\epsilon_{ab}
|\phi^b\rangle \,,
\\
&\mathfrak{S}^a_{~\alpha}|\phi^b\rangle =
c\ \epsilon^{ab}\epsilon_{\alpha\beta}
|\psi^\beta\rangle\,,
 &
\mathfrak{S}^a_{~\alpha}|\psi^\beta\rangle =
d\ \delta^\beta_\alpha |\phi^a\rangle\, ,
\label{fera0}
\eea
and for the three central extensions we have
\be \mathfrak{C} |{\cal
X}\rangle =\frac{1}2 (ad+bc)|{\cal X}\rangle\,, \quad \mathfrak{P}
|{\cal X}\rangle =ab|{\cal X}\rangle\,, \quad \mathfrak{K} |{\cal
X}\rangle =cd|{\cal X}\rangle\, ,
\ee
where the parameters $(a,b,c,d)$ are completely specified by a the
momentum $p$ of the magnon and a phase $e^{2i\xi}$.

The asymptotic states in the mirror theory are believed to transform in antisymmetric
$4Q$-dimensional representations of each  $su(2|2)$ algebra factor \cite{AF2}.  We can characterize these multiplets components, labelled by an index $i$, in terms of fundamental components in the following way \cite{bei06,CDO06}. The first $2Q$ components correspond to $\phi^{\{\alpha_1,\cdots,\alpha_{Q}\}}$
for $1\le i \le Q+1$ and to $\phi^{\{\alpha_1,\cdots,\alpha_{Q-2}\}[a,b]}$ for $Q+2\le i \le 2Q$. These
are bosonic or fermionic for $Q$ odd or even. The remaining  $2Q$ components correspond
to $\psi^{\{\alpha_1,\cdots,\alpha_{Q-1}\}a}$.
The $su(2|2)$ transformation rules on these multiplets can be obtained using
(\ref{fera0}) and  $(a,b,c,d)$  parameters
\be
a = \sqrt{\frac{g}{Q}}\eta\,,\quad
b=  \sqrt{\frac{g}{Q}}\frac{i e^{i2\xi}}{\eta}\left(\frac{z^+}{z^-}-1\right),\quad
c=- \sqrt{\frac{g}{Q}}\frac{\eta e^{-i2\xi}}{z^+}\,,\quad
d=\sqrt{\frac{g}{Q}}\frac{z^+}{i\eta}\left(1- \frac{z^-}{z^+}\right),
\label{bulkabcd}\ee
where our preferred choice for $\eta$ is
\be\label{eta}
\eta(p,e^{i2\xi}) = e^{i\xi} e^{\frac{i p}{4}}\sqrt{iz^--iz^+}\,,
\ee
and $z^\pm$ are the bound state spectral parameters
\be
e^{i p} =  \frac{z^+}{z^-}\,, \qquad
z^+ +\frac{1}{z^+}-z^- -\frac{1}{z^-}=\frac{iQ}{g}\,.
\label{sp}
\ee

The boundary scattering matrix for these bound state multiplets, in the case we are considering in this section, is obtained by demanding that an $su(1|2)^2 \subset su(2|2)^2$ symmetry be preserved by the reflection.
This $su(1|2)^2$ is the subset of the vacuum symmetries that also preserves the boundary. To do this one has to bear in mind that the action of a left boundary reflection changes  the representation labels in
the following way:
\be
{\cal R}_Q: \quad (p,e^{i2\xi}) \quad \longrightarrow \quad (-p,e^{i2\xi}e^{2ip})\,.
\ee

For $Q=1$ this was done in \cite{HMopen}. For a generic $Q$ the resulting reflection matrix
is also diagonal and entirely fixed up to an overall scalar function. For each $su(1|2)$ we obtain:
\be
{\cal R}_Q(\zeta)= {\cal R}_0(\zeta){\rm diag}(\overbrace{\vphantom{e^{\frac{i}{2}p}}1,\cdots,1}^{Q+1},
\overbrace{-1\vphantom{e^{\frac{i}{2}p}},\cdots,-1}^{Q-1},
\overbrace{-e^{\frac{i}{2}p},\cdots,-e^{\frac{i}{2}p}}^{Q},
\overbrace{e^{-\frac{i}{2}p},\cdots,e^{-\frac{i}{2}p}}^{Q})\,.
\ee
As seen in the previous section, for the boundary L\"uscher correction we need the function
\be
\chi_Q(\zeta) =
(R_Q(\zeta))_{\ j}^{i} (R_Q(-\zeta -\omega_2))^k_{\ l} {\cal C}^{jl} {\cal C}_{ik}\,.
\ee
Under $\zeta \to -\zeta -\omega_2$, the momentum dependence is unchanged: $p\to p$.
The action of the charge conjugation matrix on a bound state can be obtained from
the action of ${\cal C}$ on a fundamental magnon, which we take as \cite{AF2}
\be
\mathcal{C}_{ij} = \left(
\begin{array}{cc}
-i\epsilon_{ab} & 0
\\
0 & \epsilon_{\alpha\beta}
\end{array}
\right).
\ee
On pairs of upstairs and downstairs fundamental indices, we have
\ba
&&  {\cal C}^{1l} {\cal C}_{1k} R^{k\cdots}_{\ l\cdots} = - R^{2\cdots}_{\ 2\cdots}\,,
\qquad  {\cal C}^{2l} {\cal C}_{2k} R^{k\cdots}_{\ l\cdots} = - R^{1\cdots}_{\ 1\cdots}\,,
\\
&&{\cal C}^{3l} {\cal C}_{3k} R^{k\cdots}_{\ l\cdots} = - R^{4\cdots}_{\ 4\cdots}\,,
\qquad  {\cal C}^{4l} {\cal C}_{4k} R^{k\cdots}_{\ l\cdots} = - R^{3\cdots}_{\ 3\cdots}\,.
\ea
Therefore
\be
{\cal C}{\cal R}_Q(-\zeta-\omega_2){\cal C}^{-1}= {\cal R}_0(-\zeta-\omega_2)
{\rm diag}(\overbrace{\vphantom{e^{\frac{i}{2}p}}1,\cdots,1}^{Q+1},
\overbrace{-1\vphantom{e^{\frac{i}{2}p}},\cdots,-1}^{Q-1},
\overbrace{e^{-\frac{i}{2}p},\cdots,e^{-\frac{i}{2}p}}^{Q},
\overbrace{-e^{\frac{i}{2}p},\cdots,-e^{\frac{i}{2}p}}^{Q})
\ee
and $\chi_Q$ turns out to be exactly vanishing,
\be
\chi_Q(\zeta) = {\cal R}_0^2(\zeta){\cal R}_0^2(-\zeta-\omega_2){\rm tr}\left(
{\rm diag}(\overbrace{1,\cdots,1}^{2Q},\overbrace{-1,\cdots,-1}^{2Q})\right)^2 = 0\,.
\ee
Thus, the vacuum state energy remains vanishing for any finite length $J$.

\section{$Z=0$ brane}
\label{z=0}
A more interesting situation to consider is a $Z=0$ maximal giant graviton
with an open string, whose ground state carries angular momentum
along the $Z$ direction. Now, the dual conformal field theory operator looks like
\begin{equation}
\cO_{Z}({\cal X}_l Z^J {\cal X}_r)
=\epsilon^{i_1,\cdots, i_N}_{j_1,\cdots ,j_N}Z^{j_1}_{i_1} \cdots
Z^{j_{N-1}}_{i_{N-1}}({\cal X}_l Z^J
{\cal X}_r)^{j_N}_{i_N}\label{DefOz}\,.
\end{equation}
In this case there are some boundary degrees of freedom ${\cal X}_l$ and ${\cal X}_r$ attached to the ends
of the open chain. We will focus on the case in which ${\cal X}_l$ and ${\cal X}_r$
transform in the fundamental representation\footnote{Other possibilities exist if one adds fundamental matter to the gauge theory, corresponding to probe D-branes \cite{CY}.} under both copies of $su(2|2)^2$. The central charges of these boundary
degrees of freedom are non-trivial and such that the total energy of the ground state (\ref{DefOz}) is
\cite{HMopen}
\be
{\cal E}_0 = 2\sqrt{1+4g^2}\,.
\label{vacz0}
\ee
This expression is expected to be exact to all orders in $g^2$ only in the limit $J\to\infty$.
For any finite length vacuum, the energy (\ref{vacz0}) is valid only up to certain finite order in the weak coupling expansion. Indeed, the analogue of the closed-chain ``wrapping effects'' here is when the
range of the interaction allows the boundary degrees of freedom to perceive each other. This leads one to expect
that the leading finite size correction should occur at $g^{2J+2}$. Later we will present explicit
computations showing that this is so. Interestingly, that will mean that the leading contribution
comes from a term with $e^{-J\tilde{E}}$ rather than $e^{-2J\tilde{E}}$. This is characteristic
of a theory with one-particle interactions at the boundaries, and should manifest itself as a
double-pole in the function $\chi(q)$.

In what follows, we will use the boundary L\"uscher correction presented in section \ref{BSBTBA} to compute the leading finite size correction to the vacuum energy (\ref{vacz0}), in the weak coupling limit and for {\it any} $J$. Again, we need to compute the boundary reflection matrices for all asymptotic states in the mirror theory.

For the fundamental boundary degree of freedom we use the following parameters,
\be
a_B = \sqrt{g}\eta_B\,,\quad
b_B = -\sqrt{g}\frac{i e^{i2\xi}}{\eta_B},\quad
c_B =-\sqrt{g}\frac{\eta_B e^{-i2\xi}}{x_B}\,,\quad
d_B =\sqrt{g}\frac{x_B}{i\eta_B}\,,
\label{bndryabcd}\ee
where our preferred choice for $\eta_B$ is
\be\label{etaB}
\eta_B(e^{i2\xi}) = e^{i\xi}\frac{x_B}i\,,   \qquad x_B+\frac{1}{x_B} =\frac{i}{g}\,.
\ee
Therefore, bulk and boundary magnons are specified by $(p,e^{i2\xi})\times(e^{i2\xi}e^{ip})_B$.
The action of a boundary reflection changes these labels in
the following way \cite{HMopen}:
for a left boundary,
\be  (-e^{2i\xi})_B \times (p,e^{2i\xi}) \to (-e^{2i\xi} e^{2ip})_B \times (-p,e^{2i\xi} e^{2ip}) \,, \label{leftlabels}\ee
while for a right boundary
\be  (p,e^{2i\xi}) \times (e^{2i\xi} e^{ip} )_B \to (-p, e^{2i\xi}) \times (e^{2i\xi} e^{-ip})_B \label{rightlabels}.
\ee

The reflection matrix of $Q=1$ bulk magnons was obtained in \cite{HMopen} by imposing the requirement that the full $su(2|2)^2$ symmetry be preserved. We have extended this for generic $Q$ bulk magnons and again the boundary reflection matrix is obtained, up to an overall scalar function. We present the details of this derivation in the appendix \ref{app2}.

From now on, we will focus on the leading finite size correction for an operator
\begin{equation}
\cO_{Z}(YZ^JY) =\epsilon^{i_1,\cdots, i_N}_{j_1,\cdots ,j_N}Z^{j_1}_{i_1} \cdots
Z^{j_{N-1}}_{i_{N-1}}(Y Z^J
Y)^{j_N}_{i_N}\label{DefOz2}\,.
\end{equation}
By taking the boundary impurities to be $Y$, the whole operator is in a $su(2)$
closed sub-sector, and several explicit weak coupling computations can be made to compare against the
finite size corrections obtained from the boundary L\"uscher correction.

We need to compute $\chi_Q(\zeta)$ and evaluate it in the weak coupling limit. In order to perform  this computation, we will split $\chi_Q(\zeta)$ into scalar and matrix part factors
\be
\chi_Q(\zeta) =
 {\cal R}_Q^{sl(2)}(\zeta){\cal R}_Q^{sl(2)}(-\zeta-\omega_2)
\left(({\cal R}_Q(\zeta))^{1,i}_{1,j}
({\cal R}_Q(-\zeta -\omega_2))^{1,k}_{1,l} {\cal C}^{jl} {\cal C}_{ik}\right)^2\,,
\ee
where ${\cal R}_Q^{sl(2)}(\zeta)$ stands for the scattering factor of a $sl(2)$ bound state,
which can be obtained by standard fusion rules. In the remaining factor, $({\cal R}_Q(\zeta))^{1,i}_{1,j}$
is the reflection of an antisymmetric $Q$-bound state in which the component
${\cal R}^{3,\{3,\cdots 3\}}_{3,\{3,\cdots 3\}}$ has been set to 1.

\subsection{Matrix part factor}
For the matrix part, only certain diagonal components will contribute.
In particular, we need
\be
({\cal R}_Q(\zeta))^{1,i}_{1,i}  = \left \{
\begin{array}{c c}
a_{5,5}(-\zeta) & 1\le i \le Q+1
\\
2 a_{8,8}(-\zeta) & Q+2 \le i \le 2Q
\\
a_{9,9}(-\zeta) & 2Q+1\le i \le 3Q
\\
\frac{a_{3,3}(-\zeta)+a_{9,9}(-\zeta)}2 & 3Q+1\le i \le 4Q
\end{array}
\right.
\ee
where $a_{i,j}(-\zeta)=a_{i,j}(-z^-,-z^+)$ for the functions displayed in appendix \ref{app2}.
Then,
\ba
\!({\cal R}_Q(\zeta))^{1,i}_{1,j}{\cal C}_{ik}
({\cal R}_Q(-\zeta -\omega_2))^{1,k}_{1,l} {\cal C}^{lj} \!\!\!&=&\!\!\!
(Q+1) a_{5,5}(-\zeta) a_{5,5}(\zeta+\omega_2) + 4(Q-1)a_{8,8}(-\zeta) a_{8,8}(\zeta+\omega_2)
\nn\\
&& +\frac{Q}{2}a_{9,9}(\zeta)(a_{3,3}(-\zeta-\omega_2) +a_{9,9}(-\zeta-\omega_2))
\nn\\
&&
+\frac{Q}{2}(a_{3,3}(\zeta) +a_{9,9}(\zeta) )a_{9,9}(-\zeta-\omega_2)
\nn\\
&=&\!\!\!\!\frac{2Q(z^++z^-)(x_B^2-1)(x_B^2(z^-)^2+(z^+)^2)}
{(z^+-z^-)(x_B-z^+)(x_B-z^-)(1+x_Bz^+)(1+x_Bz^-)}
\label{exactmp}
\ea
The coupling dependence of the spectral parameters of the mirror bound state
and of the boundary parameter $x_B$ are
\be
z^{\pm} = \frac1{4g}\left(\sqrt{1+\frac{16g^2}{Q^2+q^2}}\pm1\right)\left(q+iQ\right)
\,,\qquad
x_B =\frac{i}{2g}\left(1+\sqrt{1+4g^2}\right).
\label{zpm}
\ee
Using this, the weak coupling limit for the matrix part
is\footnote{This also accounts for the $Q<3$ contributions, whose reflection matrices have to be computed separately. They were found in \cite{HMopen} ($Q=1$) and \cite{CY} ($Q=2$).}
\be
 \left(({\cal R}_Q(\zeta))^{1,i}_{1,j}{\cal C}_{ik}
({\cal R}_Q(-\zeta -\omega_2))^{1,k}_{1,l} {\cal C}^{jl} \right)^2
=  \frac{4Q^2(q^2+Q^2)^2}{\left((Q-2)^2+q^2\right)^2} +{\cal O}(g^2)\,.
\ee

\subsection{Scalar factor}
\begin{figure}
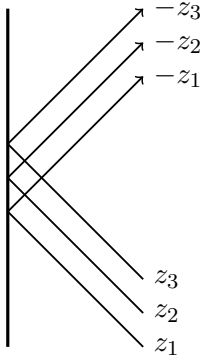

\begin{center}
\btp \draw[very thick] (0,-2) -- (0,3);

\draw[->] (2,-1.5) node [right] {$z_{2}$}
                    -- ++(-2,2) -- ++(2,2) node [right] {$-z_{2}$};
\draw[->] (2,-1) node [right] {$z_{3}$}
                    -- ++(-2,2) -- ++(2,2) node [right] {$-z_{3}$};

\draw[->] (2,-2 ) node [right] {$z_{1}$}
                    -- ++(-2,2) -- ++(2,2) node [right] {$-z_{1}$};
\etp

\end{center}
\caption{Sketch of a bound state reflection, in terms of elementary constituents scatterings.} \label{fig1}
\end{figure}

We still have to fix the scalar factor in the reflection matrices of
anti-symmetric representation magnons. We can do this by fixing the
boundary scattering factor of a $sl(2)$ bound state using fusion rules,
in terms of the scattering factors of elementary $Q=1$ constituents.
If we use $(z^+,z^-)$ for the bound state spectral parameters, the elementary
constituents $(z_1^+,z_1^-), \dots, (z_Q^+,z_Q^-)$ can be taken such that
\be
z_1^- = z^-\,, \quad z_1^+ = z_2^-\,, \quad \cdots\quad  z_Q^+ = z^+\,.
\ee
We will need boundary and bulk scattering factors involving the elementary $Q=1$ constituents,
as it is sketched in fig. \ref{fig1}.

For a $Q=1$ magnon, the scalar factor is obtained by imposing crossing symmetry \cite{ABR},
\be
{\cal R}^{sl(2)}(x) = {\cal R}_{0}^2(x) \sigma(x,-x)\sigma(x_1,-x)^2\sigma(x_2,-x)^2\,,
\label{rsl2}
\ee
where
\begin{equation}
{\cal R}_{0}^2(x) =
-\left(\frac{x^-}{x^+}\right)^2
\left(\frac{x_B-x^-}{x_B+x^+}\right)\left(\frac{x_B+\frac{1}{x^+}}{x_B-\frac{1}{x^-}}\right)
 \left(\frac{x_B+x^-}{x_B-x^+}\right)
\left(\frac{x_B +\frac{1}{x^-}}{x_B-\frac{1}{ x^+}}\right),
\end{equation}
and $\sigma(x,y)$ is the dressing factor of the bulk S-matrix, while $x_1$ and $x_2$ refer to
fundamental magnons with spectral parameters $(x_1^+=x_B,x_1^-=i)$  and $(x_2^+=i,x_2^-=-x_B)$.
We use a $sl(2)$ superscript to indicate that this is the factor when the matrix is normalized
to 1 for the 3-3 reflection.

Therefore, the scalar factor for the antisymmetric representation reflection
matrix is\footnote{To be used with the matrix that is normalized
to 1 for the  3-$\!\{3,\cdots,3\}$ reflection.}
\be
{\cal R}_Q^{sl(2)}(z) = \prod_{i=1}^{Q}{\cal R}^{sl(2)}(z_i) \prod_{j<k}^{Q}{\cal S}^{sl(2)}(-z_j,z_k)\,,
\label{fuR}
\ee
where ${\cal R}^{sl(2)}(z)$ is defined in (\ref{rsl2}) and
\be
{\cal S}^{sl(2)}(x,y)=\frac{{\cal S}^{2}_0(x,y)}{\sigma^2(x,y)}\,, \qquad {\cal S}^{2}_0(x,y)=\frac{(x^+-y^-)(1-\frac{1}{x^-y^+})}{(x^--y^+)(1-\frac{1}{x^+y^-})}\,.
\label{Ssl2}
\ee

A few comments are in order about this scalar factor. Firstly, the number of elementary scalar factors ($\tfrac{Q^2+Q}{2}$)
is larger  than in the case of a bulk scattering fusion rule ($Q$). Secondly,
some of the dressing factors appearing in (\ref{fuR}), when evaluated for magnons on-shell in the mirror theory,
are not going to be 1 at leading order in the weak coupling expansion. Therefore, it would be  difficult to know, even at leading order, the analytic continuation of an individual scalar factor ${\cal R}_Q^{sl(2)}(z)$.
However, what is needed for the finite size correction is
${\cal R}_Q^{sl(2)}(z(\zeta)) \cdot {\cal R}_Q^{sl(2)}(z(-\zeta-\omega_2))$. Fortunately, major simplifications occur for this particular product.

Contributions to ${\cal R}_Q^{sl(2)}(\zeta)\cdot{\cal R}_Q^{sl(2)}(-\zeta-\omega_2)$ can be conveniently split into three factors,
\be
\chi_0^{I}=\prod_{j<k} {\cal S}_0^2(-\zeta_j,\zeta_k){\cal S}_0^2(\zeta_j+\omega_2,-\zeta_k-\omega_2)\,,
\label{part1}
\ee
\be
\chi_0^{II}=\prod_{i}\sigma(\zeta_i,-\zeta_i)\sigma(-\zeta_i-\omega_2,\zeta_i+\omega_2)
\prod_{j<k}  \sigma^2(\zeta_k,-\zeta_j)\sigma^2(-\zeta_k-\omega_2,\zeta_j+\omega_2)   \,,
\label{part2}
\ee
and
\be
\chi_0^{III}=\prod_{i} {\cal R}_{0}^2(\zeta_i){\cal R}_{0}^2(-\zeta_i-\omega_2)
\sigma(x_1,-\zeta_i)^2\sigma(x_2,-\zeta_i)^2
\sigma(x_1,\zeta_i+\omega_2)^2\sigma(x_2,\zeta_i+\omega_2)^2 \,.
\label{part3}
\ee
The first simplification takes place in $\chi_0^{I}$, which straightforwardly becomes
\be
\chi_0^{I}=\left(\frac{z^+}{z^-}\right)^{2-2Q} \,.
\ee

The dressing functions appearing in (\ref{part2}) are not 1 in the weak coupling
limit, for magnons with mirror theory kinematics. However, the products appearing in (\ref{part2})
can be exactly computed using the following relation obtained from the crossing properties
of the bulk dressing factor,
\be
\sigma(\zeta_i,-\zeta_j) \sigma(-\zeta_i -\omega_2,\zeta_j+\omega_2) = \frac{{z_i}^+}{{z_i}^-}\frac{{z_j}^+}{{z_j}^-} \frac{f(\zeta_j,-\zeta_i)}{f(\zeta_j,-\zeta_i-\omega_2)}\,,
\ee
where the function $f(\zeta_i,\zeta_j)$ is defined by eq. (\ref{Defffunction}) in appendix
(\ref{crossbulk}). Then,
\ba
\chi_0^{II}&\!\!=\!\!&\left(\frac{z^+}{z^-}\right)^{2Q} \prod_{i} \frac{f(\zeta_i,-\zeta_i)}{f(\zeta_i,-\zeta_i-\omega_2)}
\prod_{j<k} \frac{f(\zeta_j,-\zeta_k)^2}{f(\zeta_j,-\zeta_k-\omega_2)^2}
\nn\\
&\!\!=\!\!& \left(\frac{z^+}{z^-}\right)^{2Q} \frac{4(1+z^-z^+)^2}{(z^++\frac{1}{z^+})(z^-+\frac{1}{z^-})(z^-+z^+)^2}\,.
\ea

For the final factor $\chi_0^{III}$, it is not possible to exactly simplify the dressing factors. Nonetheless, it will be convenient to use the following relation
\be
\sigma(x_1,-\zeta)\sigma(x_2,-\zeta)=-f(\zeta,x_1)f(\zeta,x_2)\sigma(x_1,\zeta+\omega_2)\sigma(x_2,\zeta+\omega_2)
\label{rela}
\ee
which is obtained using crossing and parity symmetry (and also the fact that
$\sigma(\zeta,x_1)\sigma(\zeta,x_2)=\sigma(\zeta,-x_1)\sigma(\zeta,-x_2)$).

We will use this relation and the fact that some of the dressing functions
$\sigma$ are indeed 1 at leading order in the weak coupling limit. In general, this
will depend on how the spectral parameters of the elementary constituents are taken.
We will adopt the same choice as in \cite{BJ},
\ba
&&z_1^- = z^-\,, \qquad z_k^- = z_{k-1}^+\,, \qquad  z_Q^+ = z^+\,,
\\
&&z_k^+ = \frac{1}{2}\left(z_k^-+\frac{1}{z_k^-}+\frac{i}{g}+\sqrt{\left(z_k^-+\frac{1}{z_k^-}+\frac{i}{g}\right)^2-4}\right).
\ea
Using (\ref{zpm}) and taking the weak coupling limit, one finds that (for $q>0$) all spectral parameters
of the elementary constituent are order $\tfrac1g$, except for $z_1^-$ which is order $g$.
Because of these leading behaviours, when using the perturbative expansion of the bulk dressing
factors \cite{BHL,BES}\footnote{This might seem a naive attempt to compute the analytical continuation of the dressing phase. However, it appears to be correct for the closed string computations \cite{BJ,BJL}, and we will assume it is also for the open string case we are considering here. It would be interesting to analyze the dressing phase using the integral representation of \cite{DHM} as recently done in \cite{Arutyunov:2009kf}.}, one obtains
\be
\sigma(x_1,-\zeta_i)\sigma(x_2,-\zeta_i) = 1+{\cal O}(g^6)\,, \qquad{\rm for}\quad i\ge 2
\ee
and then, using (\ref{rela}),
\be
\sigma(x_1,-\zeta_i)\sigma(x_2,-\zeta_i)\sigma(x_1,\zeta_i+\omega_2)\sigma(x_2,\zeta_i+\omega_2)
= -\frac{1}{f(\zeta_i,x_1)f(\zeta_i,x_2)}+{\cal O}(g^6)\,, \qquad{\rm for}\quad i\ge 2.
\ee
By similar arguments one can conclude that
\be
\sigma(x_1,-\zeta_1)\sigma(x_2,-\zeta_1)\sigma(x_1,\zeta_1+\omega_2)\sigma(x_2,\zeta_1+\omega_2)
= 1+{\cal O}(g^2)\,.
\label{z1sigmas}
\ee
Therefore, to capture the very leading order we can approximate
\ba
\chi_0^{III} &\!\!\sim \!\!&{\cal R}_{0}^2(\zeta_1){\cal R}_{0}^2(-\zeta_1-\omega_2)\prod_{i=2}^{Q} \frac{{\cal R}_{0}^2(\zeta_i){\cal R}_{0}^2(-\zeta_i-\omega_2)}{f(\zeta_i,x_1)^2 f(\zeta_i,x_2)^2}
\nn\\
&\!\! =\!\!& \frac{16}{((Q+2)^2+q^2)^2}\left(\frac{4g^2}{Q^2+q^2}\right)^4+ {\cal O}(g^{10})
 \,.
\label{part3apr}
\ea
Finally, for the total scalar factor we obtain
\be
{\cal R}_Q^{sl(2)}(\zeta){\cal R}_Q^{sl(2)}(-\zeta-\omega_2) =
\frac{256 q^2}{(Q^2+q^2)((Q+2)^2+q^2)^2}\left(\frac{4g^2}{Q^2+q^2}\right)^4+ {\cal O}(g^{10})
 \,.
\ee
~

\subsection{Leading finite size correction}
As discussed in section \ref{BSBTBA}, by the leading finite size correction we mean the leading weak coupling correction to the vacuum state energy. This is obtained from the boundary L\"uscher correction, which for the pseudo-energy uses the approximation,
\be
e^{-\epsilon_Q^{(0)}(q)} = e^{-2L \tilde{E}_Q(q)} = \left(\frac{4g^2}{Q^2+q^2}\right)^{2L}
+{\cal O}(g^{4L +2}).
\ee
We will therefore need to relate $L$ and $J$ according to the (somewhat awkward-seeming) $L = J -1$.\footnote{This is related to our choice of the function $\eta_B$ -- see footnote \ref{etabnote}.}

Therefore, for an operator $\cO_{Z}(YZ^JY)$
\be
\chi_Q(q)e^{-\epsilon_Q^{(0)}(q)} = \frac{1024Q^2 q^2 (q^2+Q^2)}{\left((Q-2)^2+q^2\right)^2\left((Q+2)^2+q^2\right)^2} \left(\frac{4g^2}{Q^2+q^2}\right)^{2J+2}.
\ee
We immediately recognize that, as expected, there is a double pole as $q\to0$. It is interesting to note this is entirely coming from the $Q=2$ contribution,
\be
\chi_2(q)e^{-\epsilon_2^{(0)}(q)} = \frac{4096 (4+q^2)}{q^2(16+q^2)^2} \left(\frac{4g^2}{4+q^2}\right)^{2J+2}.
\ee
Thus, for $q\to 0$,
\be
\chi_2(q)e^{-\epsilon_2^{(0)}(q)} \sim \frac{64 g^{4J+4}}{q^2}\,,
\ee
which implies, using (\ref{btbaene0bb}), that
\be
{\cal E}_0(J)-{\cal E}_0(\infty) = -4g^{2J+2} +{\cal O}(g^{2J+4})\,.
\label{leading}
\ee
Let us recall that
\be
{\cal E}_0(\infty) = 2\sqrt{1+4g^2} = 2+4g^2-4g^4+8g^6-20g^8+\cdots
\ee
For the shortest vacuum states, the leading finite size correction coming from the boundary L\"uscher correction can be compared with accessible computations at weak coupling. The above L\"uscher correction predicts, for example:
\ba
&&{\cal E}_0(0) = 2 + 0 g^2 + {\cal O}(g^4) \hspace{3.92cm}{\rm for}\quad   \cO_{Z}(YY)
\label{tba1}
\\
&&{\cal E}_0(1) = 2+4g^2-8g^4 + {\cal O}(g^6)\hspace{2.9cm} {\rm for}\quad   \cO_{Z}(YZY)
\label{tba2}
\\
&&{\cal E}_0(2) = 2+4g^2-4g^4+4g^6+ {\cal O}(g^8)\hspace{1.85cm} {\rm for}\quad    \cO_{Z}(YZ^2Y)
\label{tba3}
\\
&&{\cal E}_0(3) = 2+4g^2-4g^4+8g^6-24g^8 + {\cal O}(g^{10})\hspace{.5cm}  {\rm for}\quad    \cO_{Z}(YZ^3Y).
\label{tba4}
\ea
Some 1, 2, 3 and 4-loop gauge theory perturbative computations can be made to check
(\ref{tba1}), (\ref{tba2}), (\ref{tba3}) and (\ref{tba4}) respectively. We now proceed to perform them.

\subsection{Explicit perturbative calculations}
To test the results of the previous section we need to compute, explicitly on the gauge theory side, the scale dimension of the operators $\cO_{Z}(YZ^JY)$. These operators belong to a closed $su(2)$ sector, where the full non-planar dilatation operator is known up to two-loop order \cite{BKS},
\begin{equation}
\label{dila} D = D_0 + g^2 D_1+ g^4 D_2+{\cal O}(g^6)\,,
\end{equation}
where
\ba
D_0 \!\!&=&\!\! {\rm tr}(Z\partial_Z+Y\partial_{Y})\,, \nn
\\
D_1 \!\!&=&\!\! -\frac{2}N :{\rm tr}([Z,Y][\partial_Z,\partial_Y]):\,,
\\
\label{d2l}
D_2 \!\!&=&\!\! -\frac{2}{N^2} :{\rm tr}([[Z,Y],\partial_Z ][[\partial_Z,\partial_Y],Z]):
-\frac{2}{N^2} :{\rm tr}([[Z,Y],\partial_Y ][[\partial_Z,\partial_Y],Y]):
-2 D_1 \,.\nn
\ea

Let us first consider the action of $D_1$ on some operator $\cO_{Z}(YZ^JY)$. We have essentially
two distinct possibilities:
({\it i}) $\partial_Z$ acting on a $Z$ of the ``determinant'' (giant graviton)
or
({\it ii}) $\partial_Z$ acting on a $Z$ of the ground state.

After a careful inspection, one sees that the first possibility gives only terms which are sub-leading in the large $N$ limit. The second one gives the leading large $N$ limit terms  and to compute them, one needs to use the following the property:
\be
(\partial_{X_2}\partial_{X_1})^a_b(X_1X_2)^c_{d} = N\delta^{a}_d\delta^c_b\,.
\ee
Firstly, this kind of contribution is possible only for $L\ge 1$, because two different neighbouring scalar fields are needed somewhere in the {\it word} ${\cal W}$ defining the ground state. Secondly, these will sometimes retrieve the same original operator $\cO_{Z}(YZ^JY)$ and sometimes an operator like $\cO_{Z}(ZYZ^{J-1}Y)$. For the practical purpose, any operator $\cO_{Z}({\cal W})$, for which ${\cal W}$ begins or ends with a scalar field $Z$ should be taken as sub-leading contribution in the large $N$ limit. Such operators can be exactly re-written as a determinant times a single trace \cite{BCV1}, and the mixing with these is sub-leading in the large $N$ limit.

In conclusion, at 1-loop one obtains
\be
{\cal E}_0{(J)} =
\left\{
\begin{array} {ll}
2 +{\cal O}(g^4)\,,& {\rm if\ } J=0
 \\
2 + 4g^2  +{\cal O}(g^4)\,,& {\rm if\ } J\geq1
\end{array}
\right.
\label{d1}
\ee

The action of $D_2$ is of course a bit more involved. To begin with, now there are
leading contributions to the anomalous dimension, when one of the $\partial_Z$ acts on the determinant. However, it is important to note that these are going to be irrelevant for the leading finite size correction of the operator $\cO_{Z}(YZY)$.
This is so because, whenever a $\partial_Z$ acts on the determinant, the leading contribution is the same for $J=1$ and $J\ge 2$.

Different contributions for $J=1$ and $J\ge 2$, and therefore, the responsible ones for the leading finite size correction of  $\cO_{Z}(YZY)$, come from the action of three consecutive $\partial$ on three consecutive scalar fields. The leading terms are obtained using now
\be
(\partial_{X_3}\partial_{X_2}\partial_{X_1})^a_b(X_1X_2 X_3)^c_{d} = N^{2}\delta^{a}_d\delta^c_b\,.
\ee

To be brief, let us just quote the answer for this 2-loop computation,
\be
{\cal E}_0{(J)} =
\left\{
\begin{array} {ll}
2+ 4g^2 - 8g^2 +{\cal O}(g^6)\,,& {\rm if\ } J=1
 \\
2 + 4g^2  - 4g^4+{\cal O}(g^6)\,,& {\rm if\ } J\geq2
\end{array}
\right.
\label{d2}
\ee

~

So far, the explicit computations (\ref{d1}) and (\ref{d2}) have confirmed the leading finite size corrected
anomalous dimensions of  $\cO_{Z}(YY)$ and $\cO_{Z}(YZY)$ obtained from the L\"uscher corrections (\ref{tba1}) and (\ref{tba2}).
One would like to proceed to higher loops. However, the non-planar $su(2)$ dilatation operator is not known.
Nonetheless, we will show that it is possible to use the 3-loop and 4-loop {\it planar} $su(2)$ dilatation operator of \cite{BMR} to compute the leading finite size corrections of $\cO_{Z}(YZ^2Y)$ and $\cO_{Z}(YZ^3Y)$ respectively.

The argument is simple and runs as follow. Given the operator $\cO_{Z}(YZ^JY)$, its leading finite size correction is expected to be $(J+1)$-loop order, because this would give the minimal range of interaction needed for the boundary impurities to perceive each other. Moreover, at that order, this interaction between the boundary impurities is possible only if $(J+2)$ consecutive $\partial$'s act on the corresponding $(J+2)$ consecutive scalar fields. That is, all the $(J+2)$ $\partial$'s that would appear in the $(J+1)$-loop non-planar $su(2)$ dilatation operator have to be used.
Of course, one should also expect contributions to the anomalous dimension when 1 or more $\partial_Z$ acts on fields $Z$ of the determinant. It is just that they will never contribute to the leading finite-size correction.

Therefore, the contribution from the maximal number of consecutive $\partial$'s acting on consecutive scalar fields
should be enough to account for the leading finite size correction. These $\partial$ terms should be the same ones that generate the $(J+1)$-loop planar $su(2)$ dilatation operator.

For example, for $\cO_{Z}(YZ^2Y)$, the action of 4 consecutive $\partial$'s can be read from ${\cal H}_3$,
the 3-loop planar $su(2)$ dilatation operator \cite{BMR} in the following way. The action of ${\cal H}_3$ in a block of 4 scalar fields $(X_1 X_2 X_3 X_4)$ is given by
\ba
 \frac{({\cal H}_3)_{1234}}{g^6} &\!\!=\!\!& 60 -\frac{104}{3}({\cal P}_{12}+{\cal P}_{23}+{\cal P}_{34}) + 4 {\cal P}_{12}{\cal P}_{34}
+12 ({\cal P}_{12}{\cal P}_{23}+{\cal P}_{23}{\cal P}_{34}+{\cal P}_{23}{\cal P}_{12}+{\cal P}_{34}{\cal P}_{23})
\nn\\
&& +4{\cal P}_{12}{\cal P}_{34}{\cal P}_{23}-4{\cal P}_{23}{\cal P}_{12}{\cal P}_{34}
-4{\cal P}_{12}{\cal P}_{23}{\cal P}_{34} -4{\cal P}_{34}{\cal P}_{23}{\cal P}_{12}\,\
\label{h1234}
\ea
where  ${\cal P}_{i,i+1}$ is the permutation between two neighboring scalar fields.
The action of (\ref{h1234}) on a block $(YZZY)$ gives
\ba
{\cal H}_3  (YZZY) &\!\!=\!\!&
g^6(60-\tfrac{104}{3}) (YZZY) +
g^6(12-\tfrac{104}{3}) (ZYZY) +
g^6(12-\tfrac{104}{3}) (YZYZ)
\nn\\
&& + 8g^6 (YYZZ) + 8g^6 (ZZYY) + 4g^6 (ZYYZ)  \nn\\
&\!\!=\!\!&     g^6 (60-\tfrac{104}{3}) (YZZY) + \cdots
\ea
We added the last line to emphasize that only the first term will be relevant for us, since all
other take to a $Z$ scalar to the boundary of the block.

This has to be compared against the action of ${\cal H}_3$ on a block $(YZ^JY)$ with $J\ge3$. This action is non trivial
on  the first and the last blocks of length 4 only,
\ba
{\cal H}_3  (YZZZ\cdots) &\!\!=\!\!&  g^6 (84-\tfrac{208}{3}) (YZZZ\cdots) + \cdots
\\
{\cal H}_3  (\cdots ZZZY) &\!\!=\!\!& g^6 (84-\tfrac{208}{3}) (\cdots ZZZY) +\cdots
\ea
where we have omitted terms in which $Z$ fields are taken to the boundary. Therefore
\be
{\cal H}_3 (YZ^JY) =\left \{
\begin{array}{lll}
g^6 (60-\tfrac{104}{3}) (YZZY) + \cdots & {\rm for} & J=2
\\
g^6 (168 -\tfrac{416}{3}) (YZ^JY)+ \cdots & {\rm for} & J\ge3
\end{array}
\label{d3}
\right.
\ee
Following our previous discussion, the difference between the two lines of
(\ref{d3}) should be the leading finite size correction of  $\cO_{Z}(YZ^2Y)$.
This difference gives $-4g^6$, in agreement with the  L\"uscher correction (\ref{tba3}).

Analogously we can compute the  leading finite size correction of  $\cO_{Z}(YZ^3Y)$ from
the 4-loop planar $su(2)$ dilatation operator \cite{BMR}. The action of ${\cal H}_4$ in a block of 5
scalar fields $(X_1 X_2 X_3 X_4 X_5)$ is now given by
\ba
 \frac{({\cal H}_4)_{12345}}{g^8} &\!\!=\!\!& -560 -4\beta_{2,3}
+(268+3\beta_{2,3}+2\epsilon_{3a})({\cal P}_{12}+{\cal P}_{23}+{\cal P}_{34}+{\cal P}_{45}) \nn
\\&&
- (42+3\beta_{2,3}+2\epsilon_{3a})({\cal P}_{12}{\cal P}_{34} +{\cal P}_{23}{\cal P}_{45})
-4  {\cal P}_{12}{\cal P}_{45}    \nn
\\
&&-(\tfrac{302}{3}+\tfrac{4}3\beta_{2,3}+\tfrac83\epsilon_{3a})({\cal P}_{12}{\cal P}_{23}+{\cal P}_{23}{\cal P}_{34}
+{\cal P}_{34}{\cal P}_{45}+{\cal P}_{23}{\cal P}_{12}+{\cal P}_{34}{\cal P}_{23}
+{\cal P}_{45}{\cal P}_{34})\nn\\
&& + (2\beta_{2,3}+2\epsilon_{3a}+i\epsilon_{3c}-2i\epsilon_{3d})({\cal P}_{12}{\cal P}_{34}{\cal P}_{23}+{\cal P}_{23}{\cal P}_{45}{\cal P}_{34})
\nn\\
&& + (2\beta_{2,3}+2\epsilon_{3a}-i\epsilon_{3c}+2i\epsilon_{3d})({\cal P}_{23}{\cal P}_{12}{\cal P}_{34}+{\cal P}_{34}{\cal P}_{23}{\cal P}_{45})
\nn\\
&& + (4-2i\epsilon_{3c})({\cal P}_{12}{\cal P}_{23}{\cal P}_{45}+{\cal P}_{12}{\cal P}_{45}{\cal P}_{34})
+ (4+2i\epsilon_{3c})({\cal P}_{12}{\cal P}_{34}{\cal P}_{45}+{\cal P}_{23}{\cal P}_{12}{\cal P}_{45})
\nn\\
&&
+(48+2\epsilon_{3a})({\cal P}_{12}{\cal P}_{23}{\cal P}_{34}+{\cal P}_{23}{\cal P}_{34}{\cal P}_{45}
+{\cal P}_{34}{\cal P}_{23}{\cal P}_{12}+{\cal P}_{45}{\cal P}_{34}{\cal P}_{23})
\nn\\
&&-(6+\beta_{2,3}+2\epsilon_{3a})({\cal P}_{23}{\cal P}_{12}{\cal P}_{34}{\cal P}_{23}+{\cal P}_{34}{\cal P}_{23}{\cal P}_{45}{\cal P}_{34})
\nn\\
&& +(18+4\epsilon_{3a}) ({\cal P}_{12}{\cal P}_{34}{\cal P}_{23}{\cal P}_{45}+{\cal P}_{23}{\cal P}_{12}{\cal P}_{45}{\cal P}_{34})
\nn\\
&& -(8+2\epsilon_{3a}+2i\epsilon_{3b}) ({\cal P}_{12}{\cal P}_{23}{\cal P}_{45}{\cal P}_{34}+{\cal P}_{12}{\cal P}_{45}{\cal P}_{34}{\cal P}_{23})
\nn\\
&& -(8+2\epsilon_{3a}-2i\epsilon_{3b}) ({\cal P}_{23}{\cal P}_{12}{\cal P}_{34}{\cal P}_{45}+{\cal P}_{34}{\cal P}_{23}{\cal P}_{12}{\cal P}_{45})
\nn\\
&& -10 ({\cal P}_{12}{\cal P}_{23}{\cal P}_{34}{\cal P}_{45}+{\cal P}_{45}{\cal P}_{34}{\cal P}_{23}{\cal P}_{12})\,,
\label{h12345}
\ea
where $\beta_{2,3}=4\zeta(3)$, $\epsilon_{3a}=-2-3\zeta(3)$, $i\epsilon_{3b}=-3-\zeta(3)$.

Acting with ${\cal H}_4$ on $(YZ^JY)$ we obtain
\be
{\cal H}_4 (YZ^JY) =\left \{
\begin{array}{lll}
-\tfrac{2}{3}g^8 (338+\beta_{2,3}+2\epsilon_{3a}) (YZZZY) + \cdots & {\rm for} & J=3
\\
-\tfrac{2}{3}g^8 (332+\beta_{2,3}+2\epsilon_{3a}) (YZ^JY)+ \cdots & {\rm for} & J\ge4
\end{array}
\label{d4}
\right.
\ee
Therefore, the leading finite size correction of  $\cO_{Z}(YZ^3Y)$, given by
the difference between the two lines of (\ref{d4}), is $-4g^8$, again in agreement with the L\"uscher correction (\ref{tba4}).

As expected, this result is independent of the actual value of $\beta_{2,3}$, which is a transcendental number. If our finite size correction is right for all $J$, one should expect that the same will hold for calculations to higher-loops (and thus for longer vacuum states),  i.e. that the final result would be again independent of the $\beta_{r,s}$ appearing in the planar $su(2)$ Hamiltonian. Indeed, if one repeats the computation for $\cO_{Z}(YZ^4Y)$ using the 5-loop planar Hamiltonian quoted in \cite{BeisertThesis}, one would still get $-4g^{10}$.  This is so, despite the fact that this 5-loop planar Hamiltonian has the wrong values for the $\beta_{r,s}$ (which were fixed so as not to break BMN scaling \cite{BMN}) and demonstrates that this method of computing the leading finite size correction is independent of the actual values of $\beta_{r,s}$ also for the 5-loop example. Perhaps, by further exploiting this fact, an arbitrary loop order calculation can be performed to reproduce the leading finite size correction $-4g^{2J+2}$ of an operator $\cO_{Z}(YZ^JY)$.

\subsection{Next to leading finite size correction}
\label{ntl}
Here we will consider the next to leading order, in the weak coupling expansion, of the finite size correction. So far we have seen that, using the L\"uscher approximation to the pseudo-energy (\ref{btba}), the leading finite size correction of the operator $\cO_{Z}(YZ^JY)$ is order $g^{2J+2}$. To go beyond the L\"uscher approximation, the actual BTBA equations for pseudo-energy would be needed and it is expected they would incorporate new finite size corrections from order $g^{4J+4}$. To complete the intermediate orders one has just to continue using the L\"uscher approximation, but keep next to leading orders in the weak coupling expansion of $\chi(q)$.

This is straightforward for the matrix part, since we know the exact expression (\ref{exactmp}) and we only need to keep an extra order in the weak coupling expansion. Moreover, we only need to keep track of the $Q=2$ term, which is the one producing the double-pole in $\chi(q)$
\be
 \left(({\cal R}_2(\zeta))^{1,i}_{1,j}{\cal C}_{ik}
({\cal R}_2(-\zeta -\omega_2))^{1,k}_{1,l} {\cal C}^{lj} \right)^2
=  \frac{16(q^2+4)^2}{q^4}+ \frac{32g^2(q^4+24q^2+64)}{q^4}+{\cal O}(g^4)
\label{mpnetxtoleadingq2}
\ee

For the scalar part factor we did not use an exact expression,  and (\ref{part3apr}) can be trusted only at leading order in the weak coupling limit. This is because we have approximated
\be
\sigma(x_1,-\zeta_1)\sigma(x_2,-\zeta_1)\sigma(x_1,\zeta_1+\omega_2)\sigma(x_2,\zeta_1+\omega_2)
= 1+{\cal O}(g^2)\,,
\label{z1sigmasagain}
\ee
We need to do better than that if we want to compute the next-to-leading weak coupling order. Again, we need it only for $Q=2$ and for $q\to0$. Using the perturbative expansion of these dressing factors, one can conclude that
\be
\lim_{q \to 0}\sigma(x_1,\zeta_1+\omega_2)\sigma(x_2,\zeta_1+\omega_2) = 1+{\cal O}(g^4)
\ee
Now, we use it with (\ref{rela}) to obtain
\ba
\lim_{q \to 0}\sigma(x_1,-\zeta_1)\sigma(x_2,-\zeta_1) & \!\!=\!\!&
-\lim_{q \to 0} f(z_1,x_1)f(z_1,x_2) + {\cal O}(g^4)
\ea
Therefore
\ba
\lim_{q \to 0} \sigma(x_1,-\zeta_1)\sigma(x_2,-\zeta_1)\sigma(x_1,\zeta_1+\omega_2)\sigma(x_2,\zeta_1+\omega_2)
 &=& - \lim_{q \to 0}f(z_1,x_1)f(z_1,x_2) + {\cal O}(g^4)\nn
\\
&=& 1+ \frac{4ig^2}{q}+{\cal O}(g^4)
\ea

Then, it turns out that for the $\chi_0^{III}$ part
\ba
&&\prod_{i=1}^{2} {\cal R}_{0}^2(z_i){\cal R}_{0}^2(-z_i-\omega_2)
\sigma(x_1,-z_i)^2\sigma(x_2,-z_i)^2
\sigma(x_1,z_i+\omega_2)^2\sigma(x_2,z_i+\omega_2)^2\nn\\
&& \sim \left(\frac{f(z_1,x_1)f(z_1,x_2)}{f(z_2,x_1)f(z_2,x_2)}\right)^2 {\cal R}_{0}^2(z_1){\cal R}_{0}^2(-z_1-\omega_2)
{\cal R}_{0}^2(z_2){\cal R}_{0}^2(-z_2-\omega_2)\nn\\
&& =     \left(\frac{z^-}{z^+}\right)^{6} \frac{(x_B-z^-)^4(1+x_Bz^+)^4}{(x_B+z^+)^2(x_B+z^-)^2(1-x_Bz^-)^2(1-x_Bz^+)^2}
\nn\\
&& =  \left(\frac{4g^2}{4+q^2}\right)^4 \left(\frac{16}{(16+q^2)^2} + \frac{64g^2(q^4-12q^2-320)}{(4+q^2)(16+q^2)^3}+\cdots\right)
\label{bit3}
\ea
and this approximation could be trusted for the first two leading order of the weak coupling expansion,
in the $q\to0$ limit.
The remaining part of the scalar factor is also under control to this order.
Gathering all the contributions we find that for $q\to0$
\be
\chi_2(q)e^{-\epsilon_2^{(0)}(q)} \sim \frac{64 g^{4J+4}}{q^2} \left(1-4g^2(J+2)\right)\,.
\ee
Using (\ref{btbaene0bb}), we conclude that
\be
{\cal E}_0(J)-{\cal E}_0(\infty) = -4g^{2J+2} + 8(J+2)g^{2J+4}+{\cal O}(g^{2J+6})\,.
\label{leading2}
\ee
This is valid for $J\ge1$. For $J=0$, the next BTBA order, which we are not taking into account,
would contribute to the same sub-leading weak coupling order.

The result (\ref{leading2}) is much harder to test in general. Indeed, for the simplest case $J=1$, one would need a 3-loop order computation. This range of interaction exceeds the length of chain and so the analysis using the planar dilation operator will not suffice. Nevertheless, we can provide some evidence that this next-to-leading finite size correction is  right by using the nice interplay between alternative points of view we described in the introduction. Recall that the operators we are considering can be analyzed both
\begin{itemize}
 \item[(i)] as $Y$ impurities in the background of scalar fields $Z$, and
 \item[(ii)] as $Z$ impurities in the background of scalar fields $Y$.
\end{itemize}
Because in both cases the boundaries are specified by a determinant of scalar fields $Z$, these two points of view are genuinely inequivalent as far as the asymptotic ($J\rightarrow \8$) Bethe ansatz is concerned. The anomalous dimension of a given operator can be computed using either asymptotic Bethe ansatz, and both answers will capture the correct, finite-volume, anomalous dimension only to a certain order in weak-coupling perturbation theory. What is interesting is that the two answers generally do not fail at the same order, so that it is possible to use the more efficient viewpoint to test the finite size correction of the less efficient  one.

Let us illustrate this with two examples. Consider first the operator $\cO_{Z}(YY)$. From the point of view (i), the asymptotic description gives for the anomalous dimension $2\sqrt{1+4g^2}-2=4g^2+\cdots$. The corresponding finite size correction was already at 1-loop and precisely equal to $-4g^2$. From this point of view we have a vanishing 1-loop anomalous dimension only after incorporating the leading finite size correction. What is remarkable is that the point of view (ii), predicts a  vanishing 1-loop anomalous dimension for the operator $\cO_{Z}(YY)$, with no need to incorporate the finite size corrections corresponding to this point of view\footnote{This is quite an exceptional example, because the finite size corrections are exactly vanishing.}

Let us turn now to the operator $\cO_{Z}(YZY)$. From the point of view (i), the anomalous dimension, after taking into account {\it leading} and {\it next-to-leading}  finite size corrections (\ref{leading2}), is given by
\be
2\sqrt{1+4g^2}-2 -4g^4+24g^6 = 4g^2 -8g^4+32g^6 + \cO(g^8)
\ee
From the point of view (ii), $\cO_{Z}(YZY)$ is seen at weak coupling as a magnon with momentum $p=\tfrac{\pi}2+\cO(g^6)$. Moreover this asymptotic description is not expected to receive finite size correction until order $g^8$. Thus, by adopting the  point of view (ii), one has that the anomalous dimension of $\cO_{Z}(YZY)$ is
\be
\sqrt{1+16g^2\sin^2(\tfrac{\pi}{4}+\cO(g^6))}-1 = 4g^2 -8g^4+32g^6 + \cO(g^8).
\ee
Therefore, for the operator $\cO_{Z}(YZY)$, the ABA result for the point of view (ii)
agrees with that of (i) with the first two leading finite size correction orders incorporated.

\section{Outlook}
In this paper we used boundary the L\"uscher correction to compute the leading (and in some cases next-to-leading in $g^2$) finite-size corrections to the anomalous dimensions of various operators. For operators of the form
\begin{equation}
\cO_{Y}(Z^J)
\equiv\epsilon^{i_1,\cdots, i_N}_{j_1,\cdots ,j_N}Y^{j_1}_{i_1} \cdots
Y^{j_{N-1}}_{i_{N-1}}(Z^J)^{j_N}_{i_N}\,,
\end{equation}
we confirmed that the anomalous dimension vanishes (in the large $N$ limit) to all orders
in the 't Hooft coupling and for any $J$.

On the other hand, for operators of the form
\begin{equation}
\cO_{Z}(YZ^JY)
\equiv\epsilon^{i_1,\cdots, i_N}_{j_1,\cdots ,j_N}Z^{j_1}_{i_1} \cdots
Z^{j_{N-1}}_{i_{N-1}}(YZ^LY)^{j_N}_{i_N}\,,
\end{equation}
we found  that  the resulting predictions for the finite-size anomalous dimensions match those obtained by direct gauge theory calculations, which we also performed.

These open-boundaries calculations, on both gauge theory and worldsheet QFT sides, do not have the intricacy of the  closed-boundaries Konishi calculation in \cite{BJ,FSSZandV}.
Our hope is  that this simplicity at the \emph{leading} L\"uscher approximation
means that direct checks of subsequent orders beyond the L\"uscher approximation  against gauge theory calculations will be possible in the foreseeable future. Such tests are important to perform because they would, for the first time, involve corrections to the pseudo-energy coming from actual BTBA equations -- and they seem unlikely to be feasible in the closed-boundaries case, where the analogous TBA order would require an 8-loop gauge theory calculation.

Generalizing this boundary TBA method, in order to compute finite size correction of excited state energies, would also be of interest. This would allow for more elaborate verifications, by further exploiting the interplay between alternative points of view we described in the section \ref{ntl}.

~

\textit{Acknowledgments.-- } We thank Anirban Basu, Patrick Dorey, Anshuman Maharana and Romuald Janik for helpful discussions. D.H.C. is funded by the Seventh Framework Programme under grant agreement number PIEF-GA-2008-220702. C.A.S.Y. is funded by the Leverhulme trust.

\appendix

\section{Crossing properties of the bulk dressing factor}
\label{crossbulk}

The bulk S-matrix $S(x_1,x_2)$ takes the following schematic form:
\begin{equation}
S_{\rm full}(x_1,x_2)=S^2_0(x_1,x_2)\left(\hat{S}_{\msu(2|2)}(x_1,x_2)\otimes
\hat{S}_{\msu'(2|2)}(x_1,x_2)\right)\,.
\label{BulkSmatrix}
\end{equation}
where the overall scalar factor $S_0(x_1,x_2)^2$ is related to {\it dressing factor}
 $\sigma(x_1,x_2)$ by
\begin{equation}
S_0(x_1,x_2)^2=
\frac{(x_1^+-x_2^-)(1-\frac{1}{x_1^-x_2^+})}{(x_1^--x_2^+)(1-\frac{1}{x_1^+x_2^-})}
\frac{1}{\sigma^2(x_1,x_2)}\,,
\label{DefS0}
\end{equation}

 Recall that crossing transformation can be implemented by shifting the rapidity along the imaginary axis
$\zeta \pm \omega_2$. To implement crossing transformation in the dressing factor consistently with unitarity, one has to define the shifts in the two arguments with opposite signs. We will be using:
\begin{eqnarray}
\sigma(\zeta_1+\omega_2 ,\zeta_2) {\sigma(\zeta_1 ,\zeta_2)} &\!\!=\!\!& \frac{x^-(\zeta_2)}{x^+(\zeta_2)}{f(x_1,x_2)}\,,
\label{cs1}
\\
\sigma(\zeta_1 ,\zeta_2-\omega_2) {\sigma(\zeta_1,\zeta_2)} &\!\!=\!\!&
\frac{x^+(\zeta_1)}{x^-(\zeta_1)}{f(x_1,x_2)}\,,
\label{cs2}
\end{eqnarray}
where the function $f(z_1,x_2)$ is given by
\begin{eqnarray}
f(x_1,x_2) &\!\!\equiv\!\!&
\frac{(x^-_1-x^+_2)(1-1/x_1^+x_2^+)}{(x^-_1-x^-_2)(1-1/x_1^+x_2^-)}
=\frac{(x^-_1-x^+_2)(1-1/x_1^-x_2^-)}{(x^+_1-x^+_2)(1-1/x_1^+x_2^-)}\,.
\label{Defffunction}
\end{eqnarray}
 The unitarity condition for the scalar factor reads
\begin{equation}
S_0(z_1,z_2) S_0(z_2,z_1)=1\,, \quad \Leftrightarrow \quad
\sigma(z_1,z_2) \sigma(z_2,z_1)=1\,.
\label{ucsig}
\end{equation}

\section{Boundary reflection matrices for $Q$-magnon bound states}
\label{app2}
In this appendix we compute the reflection matrices for a bulk bound state of $Q$ magnons from a ``$Z=0$'' boundary; that is, a boundary carrying the fundamental representation with labels as in (\ref{bndryabcd}).

As in \cite{BJ} we will first consider the case of a bulk degree of freedom in a graded-\emph{symmetric} short representation, denoted $\mathcal V_Q$. This representation can be regarded
\cite{AF3} as consisting of homogeneous polynomials of degree $Q$ in the variables
$w^1, w^2, \theta^1, \theta^2$,
where the $w$ are bosonic, the $\theta$ fermionic. The generators of $su(2|2)$ are realized as
\be \mathfrak R^a{}_b = w^a \dby{w^b}  - \half \delta^a_b w^c \dby{w^c} ,
\quad \mathfrak L^\alpha{}_\beta = \theta^\alpha \dby{\theta^\beta} - \half \delta^\alpha_\beta \theta^\gamma \dby{\theta^\gamma}\ee
\be \mathfrak Q^\alpha{}_a = a\, \theta^\alpha \dby{w^a}
      + b\, \eps_{ab} \eps^{\alpha\beta} w^b \dby{\theta^\beta},\quad
    \mathfrak S^a{}_\alpha = c \,\eps^{ab}\eps_{\alpha\beta} \theta^\beta \dby{w^b}
      + d \,w^a \dby{\theta^\alpha}   \label{diffQS}\ee
with the parameters $a,b,c,d$ as in (\ref{bulkabcd}). Likewise, the boundary states correspond to homogeneous polynomials of degree 1 in variables $w'^1,w'^2,\theta'^1,\theta'^2$ and the action of the symmetry algebra is as above with primes inserted throughout and the parameters $a,b,c,d$ taken from (\ref{bndryabcd}).

The tensor product $\mathcal V_1\otimes \mathcal V_Q$ of boundary and bulk representations decomposes into the direct sum of  $10$ irreducible components\footnote{\label{etabnote}There are 10 components for all $Q>2$. For $Q=2$ there are 9 and for $Q=1$, 6. See \cite{AF3}.} with respect to the bosonic symmetries $su(2) \oplus su(2)$.  We follow precisely the conventions of \cite{BJ} in choosing bases $v^A_1,v^A_2,\dots,v^A_{10}$ of these $su(2)\oplus su(2)$-irreps and a complete set $\Lambda_i^j$ of $su(2)\oplus su(2)$-intertwiners. For the sake of brevity, the reader is referred to \cite{BJ} for the definitions. The reflection matrix is then of the form
\be \Refl(p,\xi) = \sum a_{i,j}(p,\xi) \Lambda^j_i \ee
for some coefficient functions $a_{i,j}$. As an $su(2|2)$ representation the tensor product  $\mathcal V_1\otimes \mathcal V_Q$ is irreducible (for generic values of the parameters) and so demanding that $\Refl$ commute with the supersymmetries fixes all the ratios between the $a_{i,j}$.

Let us consider a left boundary. The parameters $\xi$ and $p$ before and after scattering are then as in (\ref{leftlabels}).
In the normalization with $a_{1,1}=1$, we find that the other coefficients are as follows.

\be a_{5, 5} = \frac{\tilde\eta_B}{\eta_B} \frac{(z^-)^2+z^+x_B}{z^+(x_B+z^+)}, \qquad \qquad a_{9, 9} =\frac{\tilde\eta_B\tilde\eta }{\eta_B\eta }\frac{z^-(-x_B+z^-)}{z^+(x_B+z^+)} \ee
\be a_{5, 6} = \sqrt{Q}\frac{\tilde\eta_B}{\eta }\frac{(z^++z^-)(z^+-z^-)}{(x_B+z^+)z^+},\qquad\qquad
 a_{6, 5} = \sqrt{Q}\frac{\tilde\eta }{\eta_B} \frac{x_B(z^++z^-)}{z^+(x_B+z^+)} \ee
\be a_{6, 6} = Q\frac{\tilde\eta }{\eta } \frac{(z^+)^2-z^-x_B}{(x_B+z^+)z^+} \ee
\be a_{10, 10} = \frac{2}{Q-1}\frac{\tilde\eta ^2}{\eta ^2} \frac{(-x_B+z^-)(z^-x_B+1)}{(x_B+z^+)(z^+x_B-1)} \ee
\be a_{7, 7} = -\frac{2}{Q}\frac{\tilde\eta }{\eta } \frac{(-x_B+z^-)(z^+-x_B(z^-)^2)}{z^-(x_B+z^+)(z^+x_B-1)} \ee
\be a_{8, 8} = \half\frac{\tilde\eta_B\tilde\eta ^2}{\eta_B\eta ^2}\frac{(x_B(z^+)^2+z^-)z^-(-x_B+z^-)}{(z^+)^2(x_B+z^+)(z^+x_B-1)}\ee
\be a_{8, 7} = -\frac{i}{\sqrt Q} \frac{\tilde\eta_B\tilde\eta ^2}{e^{2i\xi}\eta }\frac{(z^++z^-)z^-(-x_B+z^-)}{(z^+)^2(x_B+z^+)(z^+x_B-1)} \ee
\be a_{7, 8} = -\frac{i}{\sqrt Q}\frac{\tilde\eta e^{2i\xi}}{\eta_B\eta ^2} \frac{(z^++z^-)(z^+-z^-)(-x_B+z^-)x_B}{z^-(x_B+z^+)(z^+x_B-1)} \ee
\be a_{3, 2} = \frac{2i}{\sqrt Q} \frac{\tilde\eta_B\tilde\eta }{e^{2i\xi}} \frac{((z^-)^2+z^+x_B)(z^++z^-)}{(z^+)^2(x_B+z^+)(z^+x_B-1)} \ee
\be a_{2, 3} = \frac{2i}{\sqrt Q}\frac{e^{2i\xi}}{\eta \eta_B} \frac{(z^-)^2+z^+x_B)(z^++z^-)(z^+-z^-)x_B}{(x_B+z^+)(z^-)^2(z^+x_B-1)} \ee
\be a_{2, 4} = \frac{i(Q-1)}{Q}\frac{e^{2i\xi}}{\eta ^2}\frac{(z^+-z^-)^2(z^++z^-)^2x_B}{(x_B+z^+)(z^+x_B-1)(z^-)^2} \ee
\be a_{3, 4} = \frac{Q-1}{\sqrt Q}\frac{ \tilde\eta_B\tilde\eta }{\eta ^2} \frac{(x_B(z^+)^2+z^-)(z^++z^-)(z^+-z^-)}{(x_B+z^+)(z^+)^2(z^+x_B-1)} \ee
\ba a_{2, 2} &=& -\frac{2}{Q(Q+1)}  \frac{1}{(x_B+z^+)(z^-)^2(z^+x_B-1)} \Big( Q (z^-)^2 z^+ +(Q+1)x_B(z^+)^2  \\ &&\qquad {} - (Q+1)x_B(z^-)^4 - Qx_B^2(z^-)^2z^+ + (z^-)^2x_B(z^+)^2 - x_B(z^-)^2\Big)\ea
\be a_{3, 3} = \frac{\tilde\eta_B\tilde\eta }{\eta_B \eta} \frac{2x_B^2(z^+)^3+(z^+)^2z^-x_B^2+(z^-)^2x_B(z^+)^2+z^+x_Bz^-+(z^-)^2z^++2(z^-)^3}{(z^+x_B-1)(z^+)^2(x_B+z^+)}\ee
\ba a_{4, 4} &=& \frac{Q-1}{2Q} \frac{\tilde\eta ^2}{\eta ^2} \frac{1}{(x_B+z^+)(z^+)^2(z^+x_B-1)}\Big((Q-1)(z^+)^4x_B-(z^+)^3z^-x_B+(z^-)^2x_B(z^+)^2\nn\\&&\qquad {} +2Q(z^+)^2z^-+Q(z^-)^2z^++Qx_B^2(z^-)^2z^++(z^-)^3x_Bz^+-Qx_B(z^-)^2\Big)\ea
\ba a_{4, 2} &=& -\frac{i}{Q}\frac{\tilde\eta ^2}{e^{2i\xi}} \frac{(z^++z^-)}{(z^+-z^-)(z^+x_B-1)(z^+)^2 (x_B+z^+)}\Big( (z^-)^2x_B(z^+)^2  - (Q-1)x_B(z^+)^2 \nn\\ &&\qquad\qquad {} - Qx_B^2(z^-)^2z^+ - z^+ x_Bz^- - Q(z^-)^2z^+ - (z^-)^3x_Bz^+ + Qx_B(z^-)^2\Big) \ea
\ba a_{4, 3} &=&\nn \frac{1}{\sqrt Q}\frac{\tilde\eta ^2}{\eta_B\eta } \frac{(z^++z^-)}{(z^+-z^-)(z^+x_B-1)(z^+)^2(x_B+z^+)}\times \\ && \qquad \Big((Q-1)x_B^2(z^+)^3 - (z^-)^2x_B(z^+)^2  - (Q-1)(z^+)^2z^-x_B^2 \nn\\ &&\qquad {} + Qz^+x_Bz^-+Qx_B^2(z^-)^2z^++Q(z^-)^2z^++(z^-)^3x_Bz^+-Qx_B(z^-)^2\Big).\ea
Here  $\eta$ and $\tilde\eta$ are the functions in (\ref{bulkabcd}) before and after scattering respectively. They are given as functions of $p$ and
 $\xi$ by (\ref{eta}); this choice is the ``string basis'' of \cite{AFZ}.
It is perhaps not entirely clear how the corresponding boundary functions $\eta_B$ and $\tilde\eta_B$ should depend on
$\xi$ and $x_B$. We have made the choice (\ref{etaB}),\footnote{Different choices for $\eta_B$ and $\tilde\eta_B$ would eventually generate extra $\tfrac{z^-}{z^+}$ factors
which might shift the relation between $J$ and $L$.} which has the merit that the reflection matrix then becomes independent of
the phase $\xi$. This in turn makes parity symmetry manifest, in the sense that the reflection matrix for a \emph{right} boundary is
(we have verified) related to the left reflection matrix above by \be \Refl^\text{right}(p) = \Refl(-p).\ee

Of course, as in \cite{BJ}, we really want the reflection matrix for graded-\emph{antisymmetric} representations, which are
believed to be the physical bound-states of the mirror theory. This amounts to exchanging $1\leftrightarrow 3$, $2\leftrightarrow 4$,
which is effectively $a\leftrightarrow d$, $b\leftrightarrow c$ in (\ref{diffQS}). Conveniently, this is in turn achieved by
$z^\pm \mapsto -z^\mp$, which is just the parity transformation, together with $\xi\mapsto -\xi+\pi$ in $a,b,c,d$, which has no effect on $\mathcal R(p)$.

\end{document}